\documentclass[
showpacs,
floatfix,
aps,
pra,
amsmath,
twocolumn,
groupaddress,
]{revtex4-1}
\usepackage{amsmath,amssymb}
\usepackage{amscd, latexsym}
\usepackage{mathrsfs}
\usepackage{graphicx}
\usepackage{epstopdf}
\usepackage{cancel}
\usepackage{amsfonts}
\usepackage{exscale}
\usepackage{dcolumn}
\usepackage{bm}
\usepackage{color}
 \usepackage{natbib}

\usepackage{lmodern}
\usepackage[T1]{fontenc}
\usepackage[latin9]{inputenc}
\usepackage{verbatim}
\usepackage{float}
\usepackage {hyperref}

\begin{document}

\title{Nonadiabatic Dynamics of a Dissipative Two-level System}

\author{Canran Xu, Amrit Poudel, Maxim G. Vavilov}

\affiliation{Department of Physics, University of Wisconsin-Madison, Wisconsin 53706, USA }

\date{February 17, 2014}

\begin{abstract}

We study the dynamics of a two-level system described by a  slowly varying Hamiltonian and weakly coupled to the Ohmic environment.  We follow the Bloch--Redfield perturbative approach to include the effect of the environment on qubit evolution and take into account modification of the spectrum and matrix elements of qubit transitions due to time-dependence of  the Hamiltonian.  We apply this formalism to two problems.  (1) We consider a qubit, or a spin-1/2, in a rotating magnetic field.  We show that once the rotation starts, the spin has a component perpendicular to the rotation plane of the field that initially wiggles and eventually settles to the value  proportional to the product of angular rotation velocity of the field and the Berry curvature.  (2) We re-examine the Landau--Zener transition for a system coupled to environment at arbitrary temperature.  We show that as temperature increases, the thermal excitation and relaxation become leading processes responsible for transition between states of the system.  We also apply the Lindblad master equations to these two problems and compare results with those obtained from the Bloch--Redfield equations.

\end{abstract}


\maketitle

\section{Introduction}
\label{sec1}
The increasing demand for accurate control of quantum devices using high-fidelity control protocols\cite{Jirari2005,Motzoi2009,Poudel2010,Bason2011}, has stimulated interest in the study of the dynamics of  quantum systems in response to slowly varying Hamiltonian. Moreover, rapid progress in the field of adiabatic quantum computing has fueled further interest in and need for more careful analysis of the dynamics of quantum systems whose parameters vary slowly in time.\cite{Childs2001} In addition, decoherence in any real quantum system  sets a rigid constraint on the time interval during which a quantum protocol must be carried out, limiting all protocols to intermediate time intervals that are shorter than the decoherence time. At these intermediate time scales, both non-adiabatic corrections and coupling to the environment become equally important.  

The previous analysis\cite{Jirari2005,Motzoi2009,Avron2010,Avron2011} of the qubit dynamics with time-dependent Hamiltonians was based on the Lindblad master equation\cite{Lindblad76,Lindblad1973} that describes the interaction with environment in terms of dephasing and transition processes characterized by phenomenological decoherence rates. An alternative microscopic approach, formulated as a perturbative theory for a quantum system with a time-independent Hamiltonian interacting with its environment, introduces the Bloch--Redfield (BR) master equation\cite{Bloch1957,Redfield1957,Schoeller1994,Makhlin2001,Makhlin2004}.  If the Hamiltonian of the system changes in time, the BR approach has to be modified to properly account for a non-adiabatic corrections.  

In this paper we extend the BR approach to account for slow evolution of the system Hamiltonian in the presence of the environment.  The main concept of the BR theory is based on the identification of decoherence processes in terms of the matrix elements for transitions caused by environment in the eigenstate basis of the quantum system.\cite{Bloch1957,Redfield1957,Makhlin2004}  
For the Hamiltonian that varies with time, one can still use a basis defined by eigenvectors of the Hamiltonian\cite{Ao1989,Ao1991,Schoeller1994,Grifoni1998,Vitanov1999,Shevchenko2004,Whitney2005,Pokrovsky2007,Pekola2010,Saha2012}, 
where the Hamiltonian is always represented by a diagonal matrix $\tilde H =U HU^\dagger$, where the unitary transformation $U$ denotes a transition from the original basis to the eigenstate basis.  Time-dependence of $U$ produces an extra term in the time evolution of the quantum system that is effectively described by the new Hamiltonian $\tilde H - i U \partial_t U^\dagger$.  This expression is not necessarily diagonal and another basis transformation is required.  Such series of diagonalization transformations can be continued indefinitely, but for slowly changing Hamiltonian, the series can be truncated after a finite number of transformations neglecting terms of the higher order in time-derivatives of the parameters in the  Hamiltonian.  In addition to changes in the effective spectrum of the system, matrix elements representing coupling between the quantum system and its environment are also modified, resulting in a redefinition of the transition rates for the system.  

We focus our analysis on the dynamics of a two-level quantum system --- a qubit or a spin-1/2 system --- in the presence of time-dependent field, which we refer to below as the control field of the qubit.  
We study the dynamical response of the transverse magnetization to quench velocity of the control field.  The transverse magnetization measurements can provide the value of the Berry curvature of a quantum system~\cite{Avron2011,Gritsev2011} and,consequently, characterize topological properties of a ground state of the system.

Since any real qubit is always coupled to its environment, it is necessary to perform detailed analysis of the non-adiabatic dynamics of a qubit system in the presence of dissipation.  To this end, we investigate the effect of pure dephasing and energy relaxation due to the Ohmic bath on the qubit polarization. Our results indicate that the decoherence suppresses the transient wiggles of the out-of-plane qubit projection,  thereby bringing up the linear relation between the qubit response and the quench velocity. Thus, the dissipation facilitates the Berry curvature measurement based on the non-adiabatic response, proposed in Ref.~\cite{Gritsev2011}. Furthermore,  this study is also related to the measurement technique of the Berry phase in qubits, which are based on an interference experiment where the drive parameter was changed slowly~\cite{Leek2007}, see also Refs.~\cite{Whitney2004,Whitney2005,DeChiara2006} for theoretical analysis of the influence of environment on the Berry phase. 

We also apply the modified 
BR equation to the Landau--Zener (LZ) problem\cite{Landau1932,Zener1932,Stueckelberg1932,Majorana1932} in a qubit coupled to environment at arbitrary temperature.  The LZ problem in a quantum system coupled to its environment has attracted significant interest recently, where the environment was considered either as a source of classical noise\cite{Pokrovsky2003,Kenmoe2013}, or quantum fluctuations that cause transitions between qubit states \cite{Wubs2006,Wubs2005,Saito2007,Pokrovsky2007,Orth2010,Whitney2011,Sun2012}, or pure dephasing\cite{Shimshoni1993,Avron2010}.  More recently, the LZ interferometry has attracted a growing interest\cite{Rudner2008,Ashhab2007,Shevchenko2010,Gasparinetti2011,Ganeshan2013,Forster2013,Tan2014}.
Here we focus on the role of quantum fluctuations in the environment that cause transitions between the eigenstates of the qubit in the LZ problem.  We argue that during the LZ transition, the matrix elements of the coupling between the qubit and its environment must be considered in the basis of eigenstates of the full qubit Hamiltonian and therefore, the matrix elements acquire an explicit time dependence due to rotation of the eigenstate basis in addition to straightforward dependence on the energy difference between the eigenstates.  This treatment modifies the previous results of Refs.~\cite{Wubs2006,Wubs2005,Saito2007} and generalizes the results of Refs.~\cite{Pokrovsky2007,Orth2010}, where a similar basis transformation was naturally included in the calculations.  We disregard the effect of the Lamb--Stark shift on the qubit spectrum due to coupling to the environment, considered in Ref.~\cite{Whitney2011}, since this can be included in the redefined control field of the qubit.  We focus solely on the transition effects due to non-unitary evolution of the qubit density matrix.  We  consider the quantum fluctuations of the environment that are fixed along the direction of the control field at very long initial and final moments of the LZ transition so that the matrix element that characterizes the transition between qubit states at long times is absent and environment produces dephasing only.  For arbitrary direction of the fluctuating field, the transition remains effective over long time and will effectively bring the qubit to the ground state for zero temperature environment. We also consider ``dephasing'' coupling\cite{Shimshoni1993,Avron2010} when the quantum fluctuations occur only in the direction parallel to the direction of the control field in the parameter space of the qubit Hamiltonian. Our result is in agreement with Ref.~\cite{Avron2010} of the same problem within Lindblad master equation, in the limit of a high-temperature environment. 

This paper is organized as follows. In section \ref{sec2}, we present a formalism of the BR equations in transformed basis for time-dependent Hamiltonians. In section \ref{sec3}, we study the evolution of a qubit whose control field rotates in a plane with a constant magnitude and consider different directions of the environmental coupling field. In section \ref{sec4}, we consider the LZ problem in the presence of zero and finite temperature environment and show that transition is dominated by thermal excitation of the qubit at finite temperatures. In section \ref{sec5}, we analyze the non-adiabatic effects within the Lindblad formalism. We end with conclusions in section \ref{sec6}.

\section{Bloch--Redfield approach to time-dependent Hamiltonians}
\label{sec2}
We consider a spin coupled to a bath of harmonic oscillators.  The full Hamiltonian 
$\hat{H}=\hat{H}_0+\hat{H}_{\rm int}+\hat{H}_{\rm env}$ is a sum of the Hamiltonian for the spin in the magnetic field $\bm{b}(t)$  
\begin{equation}
\label{eq:H0}
\hat{H}_0= - \frac{1}{2} \bm{b}(t)\cdot \hat{\bm{\sigma}},
\end{equation}
the interaction Hamiltonian of the spin with the environment~\cite{Leggett1987}
\begin{equation}
\label{eq:Hint}
\hat{H}_{\rm int}=\sum_q \lambda_q \bm{n} \cdot \hat{\bm{\sigma}}\, \frac{\hat{a}_q+\hat{a}_q^\dagger}{2}
\end{equation}
and the bath Hamiltonian 
\begin{equation}
\hat{H}_{\rm env}=\sum_q \hbar\omega_q (\hat{a}_q^\dagger \hat{a}_q+1/2).
\end{equation}
Here we assume that each environment oscillator interacts with the spin as a quantized magnetic field $\lambda_{q} (\hat{a}^{\dagger}_{q}+\hat{a}_{q} )/2$ in the common direction $\bm{n}$, $\hat{a}_{q}^{\dagger}$ and $\hat{a}_{q}$ are raising and lowering operators of the field. 

The reduced density matrix $\hat{\rho}$ of the spin is determined by tracing out environment degrees of freedom of the full density matrix $\hat{\rho}_{\rm full}$.  The full density matrix satisfies the unitary  master equation
\begin{align}
\frac{d \hat{\rho}_{\rm full} (t)}{dt}  = \frac{1}{i\,\hbar} \Big[\hat{H} (t), \hat{\rho}_{\rm full} (t)\Big]\,.
\end{align}
There are several approaches to obtain the corresponding equations for time evolution of the reduced density matrix for the qubit.  Here we consider the limit of weak coupling of a qubit to the environment, when the density matrix is defined by the BR equations,\cite{Bloch1957,Redfield1957} see also Refs.~\cite{Schoeller1994,Makhlin2004,Pokrovsky2007} where a diagrammatic  technique was developed to treat the weak coupling to environment.

The environmental effects are characterized by the spectral density function of the coupling
 $J(\epsilon) = \pi\sum_q\lambda^2_q\delta(\epsilon-\hbar \omega_q)$.
A generic spectral function has a power law dependence on energy at small energies, $J(\epsilon)\sim\epsilon^s$, and vanishes rapidly for energies above the ultraviolet cutoff $E_c$. 
Here, we consider the Ohmic ($s=1$) environment with exponential high-energy cutoff:
\begin{equation}
\label{eq:J}
J(\epsilon)= 2\pi \alpha\epsilon \exp(-\epsilon/E_c),  
\end{equation}
where the dimensionless parameter 
$\alpha$ defines the strength of coupling between the qubit and its environment and $E_c$ is the cutoff. We restrict ourself to the weak coupling limit, $\alpha \ll 1$. 

In general, the effect of weak environment on the qubit dynamics is twofold. On one hand, the qubit Hamiltonian is renormalized by the environment modes with $\epsilon<E_c$, known as the Lamb and Stark effects. On the other hand, when we integrate out the environmental degrees of freedom, we also obtain non-unitary terms in the evolution of the quantum system.  Both of these effects are accounted for by the BR equation\cite{Redfield1957,Bloch1957,Makhlin2004} for the qubit density matrix $\hat \rho (t)$. 

We first consider the case of a constant external magnetic field along $\hat z$ direction,  $\bm{b}=b \hat{\bm{z}}$. Then, the BR equation has the following form in the eigenstate basis
\begin{subequations}
\label{eq:BRall}
\begin{align}
\left(   \begin{array}{c}
    \dot \rho_{00}(t) \\ 
    \dot \rho_{11}(t) \\ 
  \end{array}\right) & =
\left(
  \begin{array}{cc}
    -\Gamma_{e} & \Gamma_{r} \\ 
    \Gamma_{e} & -\Gamma_{r} \\ 
  \end{array}
\right)
\left(   \begin{array}{c} 
    \rho_{00}(t) \\ 
    \rho_{11}(t) \\ 
  \end{array}
\right), 
\label{eq:BRdiag}\\
\dot \rho_{01}(t) & = (i \epsilon -\Gamma_{2}) \rho_{01}(t),\label{eq:BR01}\\
\dot \rho_{10}(t) & = (-i \epsilon -\Gamma_{2}) \rho_{10}(t).\label{eq:BR10}
\end{align}
\end{subequations}
We obtained the above equations within secular approximation that neglects fast oscillating terms with frequencies larger than the decoherence rates.

The equation in the matrix form, 
Eq.~\eqref{eq:BRdiag}, determines the evolution of diagonal elements of the density matrix.  
The relaxation and excitation rates, $\Gamma_r$ and $\Gamma_e$, are defined by the spectral density $J(\epsilon)$ at the energy corresponding to the energy difference between two states of the qubit:
\begin{subequations}
\label{eq:gamma_re}
\begin{align}
\Gamma_{r} & = \frac{n_{x}^{2}+n_{y}^{2}}{2\hbar}J(\epsilon)(N(\epsilon)+1), \\
\Gamma_{e} & = \frac{n_{x}^{2}+n_{y}^{2}}{2\hbar}J(\epsilon) N(\epsilon),
\end{align}
and $N(\epsilon)=1/[\exp(\epsilon/T)-1]$ is the Planck's function.  The factor $n_{x}^{2}+n_{y}^{2}$ indicates that only the component of the 
fluctuating environment field that is perpendicular to the direction of the control field $\bm{b}$ gives rise to the qubit flip processes. 

The off-diagonal elements of the density matrix are characterized by the decoherence rate $\Gamma_2$ and pure  dephasing rate $\Gamma_{\varphi}$ given by
\begin{equation}
\label{eq:gamma2}
\Gamma_{2}  =\frac{1}{2}\left(\Gamma_{r}+\Gamma_{e}\right)+\Gamma_{\varphi},
\quad \Gamma_{\varphi} =
n_{z}^{2}J_0.
\end{equation} 
\end{subequations}
The decoherence stems from two processes --- the qubit flip processes with rate $\Gamma_{r}+\Gamma_{e}$, and pure dephasing which is not responsible for energy transitions at low frequency with rate $J(\epsilon \simeq 0)\equiv J_0$.
The only source of pure dephasing is the fluctuating fields of the environment along the external field $\bm{b}$, hence the factor $\cos\theta$ in the definition of the pure dephasing term, $\Gamma_{\varphi}\propto n_{z}^{2}$.

The renormalization of the qubit Hamiltonian by the environment due to the Lamb or Stark effects are determined by the imaginary part of the environmental correlation function, as discussed in Ref.~\cite{Makhlin2004}.  Explicitly, the renormalized qubit energy $\epsilon$ is
\begin{equation}
\epsilon=b+\delta \epsilon,\quad
\delta \epsilon = -{\rm{P}}\int \frac{d\omega}{4\pi} \frac{J(\omega)\coth(\omega/2T)}{\omega-b},
\end{equation}
where $\rm{P}$ denotes the Cauchy principal value.  Below, we assume that 
the control field $\bm{b}$ already includes renormalization effects from the environment.
The goal of this paper is to investigate the
features of the qubit evolution originating 
from decoherence
characterized by rates $\Gamma_{r}$ and $\Gamma_{e}$, respectively.  The 
significance of the effect of the Lamb and Stark shifts on the evolution of the qubit was demonstrated in Ref.~\cite{Whitney2011} in the context of the LZ problem.

We note that the qubit density matrix can be defined in terms of the magnetization in $x$, $y$ and $z$ directions as
\begin{align}
\label{eq:re_mag}
\hat{\rho}(t) & =\frac{1}{2}\left(
1+\bm{m}(t)\cdot\bm{\sigma}\right).
\end{align}
Then the BR equations, Eq.~\eqref{eq:BRall}, acquire a more common form of the Bloch equations
\begin{subequations}\label{eq:Blochall}
\begin{align}
\dot{m}_z &= (\Gamma_r-\Gamma_e)-(\Gamma_r+\Gamma_e)m_z, \\
\dot{m}_x &= -i\epsilon m_y - \Gamma_2 m_x \\
\dot{m}_y &= i\epsilon m_x - \Gamma_2 m_y.
\end{align}
\end{subequations}

The above BR equations were obtained in the basis of qubit eigenstates. In case when the control field $\bm{b}(t)$ changes in time, 
we perform transformation $\hat{U}_1(t)$ of the basis that keeps the  qubit Hamiltonian diagonal.  This basis is commonly referred to as adiabatic. The corresponding  transformation has two consequences. 

The first consequence of $\hat{U}_1(t)$ transformation is that the Hamiltonian in the new basis
acquires an extra term originating from the time dependence of the transformation  $\hat{U}_1(t)$. Thus, the qubit Hamiltonian in the new basis is
\begin{align}
\label{eq:HU1}
&\hat{H}_0^{U_1} (t) = 
-\frac{\epsilon(t)}{2} \,\hat{\sigma}_z -  i\hat{U}_1(t) \dot{\hat{U}}_1^{\dagger}(t)  \,. 
\end{align}
The resulting Hamiltonian still may remain non-diagonal due to the 
Berry connection term, $ i\hat{U}_1(t) \dot{\hat{U}}_1^{\dagger}(t)$.  We can introduce a new transformation $\hat{U}_2(t)$ that diagonalizes the right hand side of \eqref{eq:HU1}, but this transformation generates a new 
term $ i\hat{U}_2(t) \dot{\hat{U}}_2^{\dagger}(t)$ and the ``diagonalization'' series of transformations $\hat{U}_n(t)$ does not stop for an arbitrary time evolution of $\bm{b}(t)$, because the Berry connection terms appearing in each consecutive diagonalization transformation acquires an extra time derivative. However, for slow time evolution, the series of transformations can be truncated by the first one or two transformations.  
Since the BR treatment of environmental effect requires anyway that the system changes  in time slower than the rates given by Eqs.~\eqref{eq:gamma_re} and \eqref{eq:gamma2} in the master equation, the truncation to a limited number of transformations $\hat{U}_n(t)$ under slow evolution of $\bm{b}(t)$ is justified. 
Also, in a special case of constant rotation of $\bm{b}(t)$ in a plane, the second transformation $\hat{U}_2$ is time-independent and transformation series stops after this second basis rotation.

The second consequence of the basis transformations is the modified interaction term in that the coupling between the qubit and its environment 
\begin{equation}
\label{eq:HintV}
\tilde{\hat{H}}_{\rm int}=\sum_q \lambda_q \bm{n}'(t) \hat{\bm{\sigma}}  \frac{\hat{a}_q+\hat{a}_q^\dagger}{2},\quad
\bm{n}'(t)  \hat{\bm{\sigma}}=\bm{n}  \hat{V} (t) \hat{\bm{\sigma}} \hat{V}^\dagger (t)
\end{equation}
is modified from the initial coupling operator  $\bm{n}\cdot\bm{\sigma}$ to the environment  field by the transformation matrix $\hat{V}(t)=\hat{U}_n(t)\dots \hat{U}_1(t)$.
This transformation changes the corresponding ``projection'' factors $n_{\{x,y,z\}}$ in  Eqs.~\eqref{eq:gamma_re} as well as the spectral weights $J(\epsilon)$.

Modification of the coupling between the qubit and its environment, introduced by Eq.~\eqref{eq:HintV}, swaps components of the fluctuating field responsible for the pure dephasing and transition processes. For example, in case of a  fixed external field $\bm{b}\| \bm{e}_{z}$, fluctuations along $\bm{e}_z$ give rise to pure dephasing and do not cause transition processes between qubit eigenstates.
However, as $\bm{b}(t)$ rotates while $\bm{n}$ remains in $\bm{e}_z$ direction, the fluctuating component along field $\bm{b}(t)$ is the only one responsible for the dephasing with the corresponding rate proportional to the spectral weight of its low-frequency fluctuations $J_0$, while
the  component of the fluctuating field perpendicular to $\bm{b}$ will produce qubit flip processes with the rate characterized by the spectral weight of fluctuating field with the energy equal to the energy of qubit flip $J(\epsilon(t))$.  The second unitary transformation further mixes matrix elements of the coupling to environment representing qubit flip processes and pure dephasing.

Below, we present explicit expressions for the rates in Eqs.~\eqref{eq:gamma_re} for two special cases of evolution of $\bm{b}(t)$ for different types of environment.
We focus on the effect of qubit flip processes due to environment and assume that $J_0=0$ in most numerical solutions.  We note that the pure dephasing produced by the low frequency noise of the environment can be successfully described in terms of fluctuations of the classical field and may also include non--Markovian time correlations that are omitted in the BR approach. Effects of classical noise were discussed in Refs.~ \cite{Kayanuma1984,Shimshoni1993,Saito2002,Pokrovsky2003,Vestgarden2008,Kenmoe2013} for the LZ transition and in Refs.~\cite{Whitney2004,Whitney2005} for Berry phase measurements.

\section{Qubit rotation in a plane}
\label{sec3}

We first consider a qubit with the Hamiltonian characterized by a time-dependent field in $x-z$ plane: $\bm{b}(t)=\Delta\{\sin\theta(t), 0, \cos\theta(t)\}$.  By definition, $\theta(t) = 0$ for $t<0$. The transformation to adiabatic basis is defined by:
\begin{equation}
\label{eq:u1}
\hat{U}_1(t)=\exp(i\hat{\sigma}_y \theta(t)/2)
 \end{equation}
 and the resulting qubit Hamiltonian has the form 
\begin{equation}
\label{eq:HU0_rot}
\hat{H}^{U_1}_0=-\frac{\Delta \hat{\sigma}_z+\dot\theta(t)\hat{\sigma}_y}{2}.
\end{equation} 
Here, the second term is responsible for the non-diagonal form of the Hamiltonian for time-dependent rotation angle $\theta(t)$ and causes the resultant field to point out of the rotation plane of $\bm{b}(t)$.  This Hamiltonian has  eigenvalues $\varepsilon_\pm=\pm\sqrt{\Delta^2+\dot \theta^2}/2$ and   eigenvectors, which are different from the vectors of the adiabatic basis. The latter two represent spin states in the ($x-z$) plane with $m_y=0$.  On the contrary,   the qubit in the ground state $|g\rangle$ of the Hamiltonian~\eqref{eq:HU0_rot} has a non-zero expectation value of the polarization $m_y$ in the direction perpendicular to the ($x-z$)  plane of the control field $\bm{b}$: 
\begin{equation}
\label{eq:mbot-}
m_y =\langle g| \hat{\sigma}_y |g\rangle = - \frac{\dot \theta}{\sqrt{\Delta^2+\dot \theta^2} }.
\end{equation}
In the limit of slow rotations, $\dot\theta(t)\ll \Delta$, this result is consistent with a more general expression that connects a generalized force $f_i=-\langle g |\partial \hat{H}(\bm{X})/\partial X_i| g \rangle $ to time-dependent parameters $\bm{X}(t)$ of the Hamiltonian through the Berry curvature $F_{ij}$ as \cite{Gritsev2011,Avron2011}
\begin{equation}
\label{bc_b}
f_i = -\langle g |\frac{\partial \hat{H}(\bm{X})}{\partial X_i} | g \rangle=  \sum_j F_{ij}\dot X_j(t).
\end{equation}
Comparing Eq.~\eqref{eq:mbot-} and Eq.~\eqref{bc_b}, we identify $f_y=m_y/2$, $\dot X= \dot \theta$ and $F_{y\theta}=1/(2\Delta)$.
Explicitly, the coefficient of the linear term in the rate of change of the magnetic field, \textit{i.e.} $\Delta\Omega$, is  the Berry curvature $1/2\Delta^2$.  Indeed, this value of the Berry curvature gives the Berry phase $\Phi=\pi$ for one full rotation of the control field in the $(x-z)$ plane after its integration over the half-sphere, $\int_{S(\bm{b})} ds/(2\Delta ^{2})  = \pi$.
This relation holds for an isolated  qubit controlled by field $\bm{b}(t)$, assuming that  $\bm{b}(t)$ is a slowly varying function of time with continuous higher derivatives. 

However, if the rotation of the control field $\bm{b}$ starts instantaneously with constant angular velocity $\dot\theta(t)=\Omega$, \textit{i.e.} $\theta(t)=\Omega t$, the rotation is equivalent to a quantum quench in the representation of Eq.~\eqref{eq:HU0_rot} from $\dot\theta =0$ to $\dot\theta=\Omega$.  The qubit that was initially in the ground state of the original time-independent Hamiltonian, $-b\hat{\sigma}_z/2$, is in the superposition of eigenstates of the  new Hamiltonian and exhibits precession around new direction of the effective field $(0, \Omega, b)$.  This precession causes oscillations of 
\begin{equation}
\label{eq:mbot}
m_y(t)= {\rm Tr}\{ \hat{\sigma}_y \hat{\rho}(t)\} 
\end{equation}
around its average value given by Eq.~\eqref{eq:mbot-}.  In this section we demonstrate that a qubit coupled to a zero-temperature environment relaxes towards the lower eigenstate of Hamiltonian \eqref{eq:HU0_rot} and for long time limit after the rotation started, the qubit state obeys Eq.~\eqref{eq:mbot-}. 

For rotation with constant angular velocity $\Omega$, the transformed Hamiltonian, Eq.~\eqref{eq:HU0_rot} is time independent and can be diagonalized by
the second basis transformation 
\begin{equation}
\label{eq:U2}
\hat{U}_2=\cos\eta/2 + i \hat{\sigma}_x \sin\eta/2,\quad \tan\eta=\Omega/\Delta .
\end{equation} 
The qubit Hamiltonian in a new basis after a full transformation $ \hat{V} (t) = \hat{U}_2 \hat{U}_1(t)$ becomes fully diagonal with time-independent eigenvalues:
\begin{equation}
\label{eq:HV}
\hat{H}_0^{V}= \hat{U}_2 \hat H_0^{U_1} \hat{U}_2^\dagger = - \frac{W}{2}\hat\sigma_z\,,
\quad 
W = \sqrt{\Delta^2+ \Omega^2}.
\end{equation}
We can apply the BR  equation for the qubit density matrix, where the rates in Eq.~\eqref{eq:BRall} are defined by the interaction term $H_{\rm int}$, Eq.~\eqref{eq:Hint}, with $\bm{\sigma} \cdot \bm{n}$ replaced by its transformation under $ \hat{V}(t)$ according to Eq.~\eqref{eq:HintV}. 
The result of the $ \hat{V}(t)$ transformation depends on the original orientation of the vector $\bm{n}$ in the qubit space.  Below, we consider three orientations of $\bm{n}$. 
We note that for the limit $\Omega\ll \Delta$ considered in this section, the shift of eigenvalues of Hamiltonian \eqref{eq:HV} and modification of the coupling to environment by the second transformation $\hat{U}_2\simeq 1$ is not significant and can be disregarded to the lowest order in $\Omega$.

\begin{figure} 
\begin{centering}
\includegraphics[width=0.9\columnwidth]{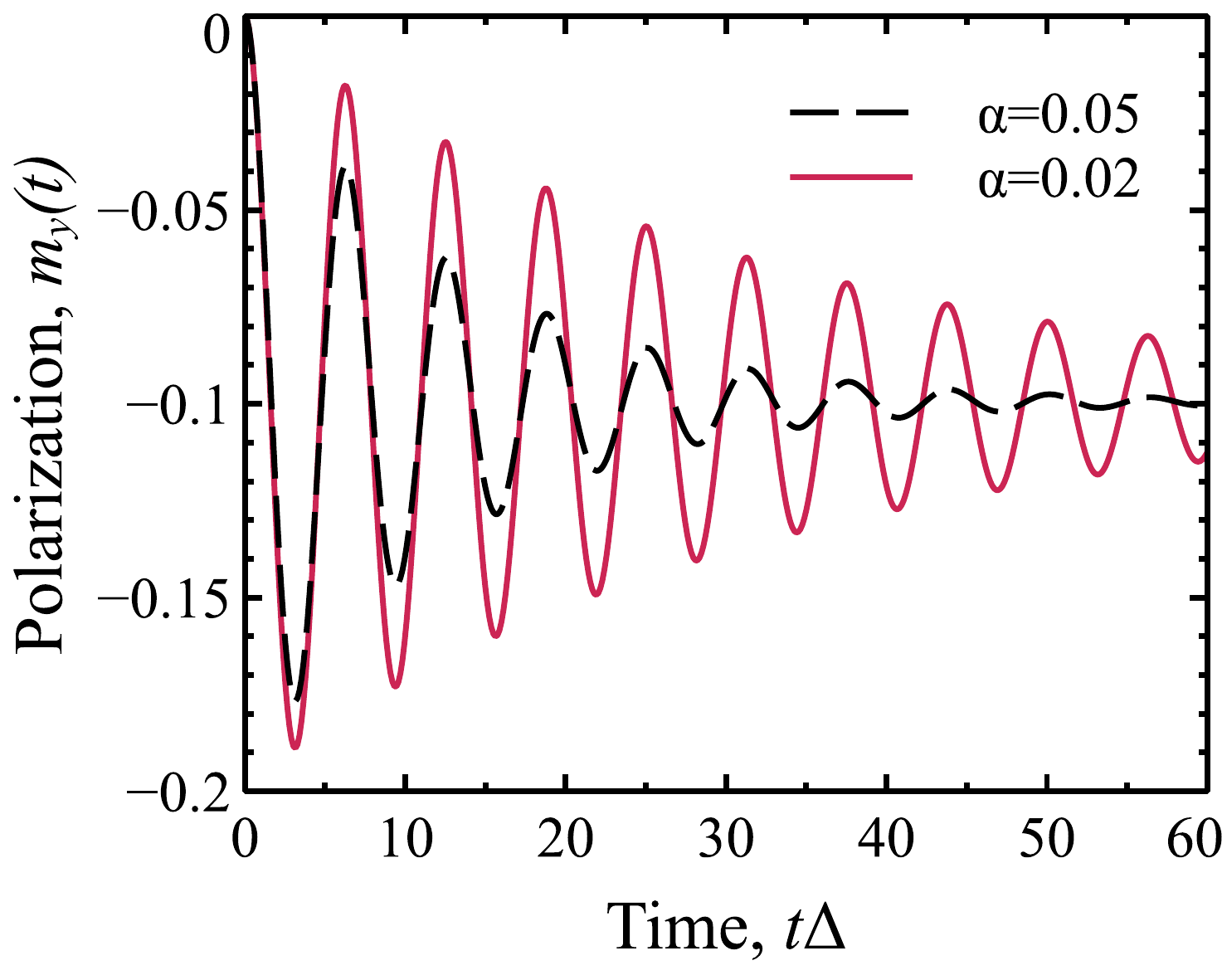}
\end{centering}
\caption{\label{fig:3A_my_vs_t_variosgammaTzero}(Color online) Time dependence of the out-of-plane polarization, $m_y(t)$, at zero temperature of environment for  $\alpha=0.02$ (solid line) and $\alpha=0.05$ for fluctuating environment field out of the plane of rotation, $\bm{n}=\hat{\bm{y}}$. The pure dephasing rate is zero, $J_0=0$. The frequency of rotation of the control field is $\Omega=0.1\Delta$.}
\end{figure}

\subsection{Environment field perpendicular to the rotation plane}

We first consider the case when the coupling between the qubit and its environment is determined by the vector  $\bm{n}=\bm{\hat {y}}$ perpendicular to the plane of rotation of the external field $\bm{b}(t)$.  For time independent Hamiltonian, this coupling causes qubit flip processes and the corresponding decoherence rates are defined by the environment spectral function at the excitation energies equal to the qubit energy splitting. For time-dependent Hamiltonian with rotating $\bm{b}(t)$, we have to write the qubit coupling operator $\bm{n}\cdot \bm{\sigma}$ in the rotated basis that diagonalizes the original Hamiltonian.  As we discussed above, the transformation is a product of two consecutive transformations. 
The first transformation, $\hat{U}_1(t)$ to the adiabatic basis does not change the coupling operator $\hat{U}_1(t)\, \bm{n} \cdot \bm{\sigma}\,\hat{U}_1^\dagger (t) =\hat{\sigma}_y$.   The second transformation results in 
\begin{equation}
\hat\Sigma_y = \hat{V}(t)\, \hat{\sigma}_y \hat{V}^\dagger(t)  = \hat{\sigma}_y \cos\eta - \hat{\sigma}_z\sin\eta.
\end{equation}
Here, the first term represents the qubit flip process, while the second term preserves the qubit orientation and causes pure dephasing. 
The corresponding rates in the BR equations are given by
\begin{subequations}\label{eq:ratesA}
\begin{align}
\Gamma_{r} & = \frac{\cos^{2}\eta}{2}J(W) [N(W)+1], \\
\Gamma_{e} & = \frac{\cos^{2}\eta}{2}J(W)  N(W),\\
\Gamma_{2} & = \frac{\Gamma_{r}+\Gamma_{e}}{2}
+
\frac{\sin^{2}\eta}{2}J_0,
\end{align}
\end{subequations}
with $W$ and $\eta$ defined by Eqs.~\eqref{eq:HV} and \eqref{eq:U2}.
The qubit dynamics is characterized by the relaxation and excitation rates proportional to the spectral function $J(W)$ of environment at energy $W$, these rates appear with factor $\cos^{2}\eta=\Delta^{2}/W^2$ and recover the case of the qubit with a time-independent Hamiltonian with $\bm{b}\bot \bm{n}$ when only environment modes in resonance with the qubit contribute to the qubit dynamics. At finite $\Omega$, however,  the pure dephasing mechanism arises after transformation $\hat{U}_{2}$ and originates from the low frequency modes of the environment with spectral density $J_0$.  The pure dephasing rate contains factor $\sin^{2}\eta=\Omega^{2}/W^2$ which is small for slow rotation with $\Omega\ll \Delta$.  

\begin{figure}
\begin{centering}
\includegraphics[width=0.9\columnwidth]{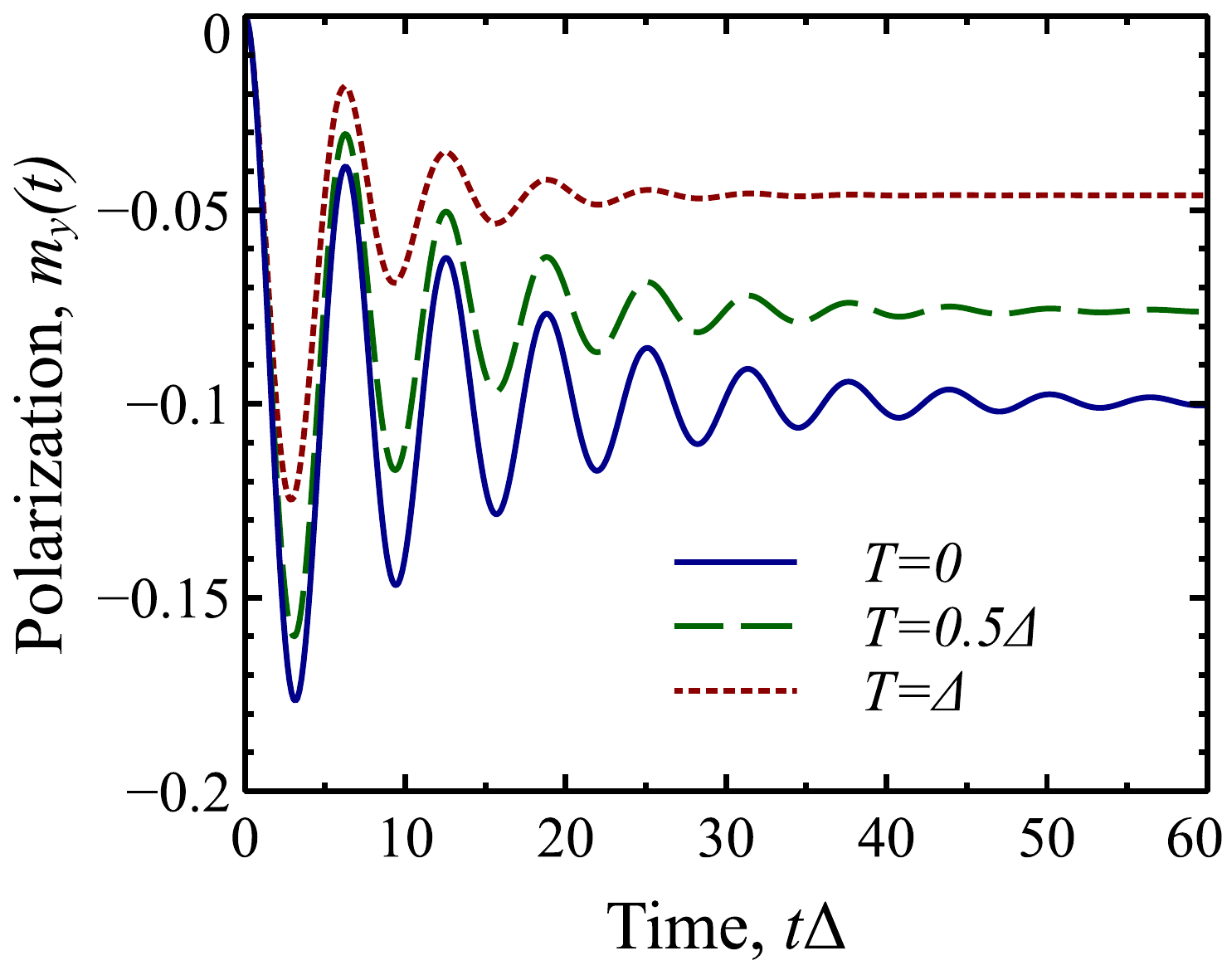}
\end{centering}
\caption{\label{fig:3A_my_vs_t_for_temp}(Color online) Time dependence of the out-of-plane polarization, $m_y(t)$, at various temperatures of environment: $T=0$ (solid line), $T=0.5\Delta$ (dashed line) and $T=\Delta$ (dotted line) for fluctuating environment field out of the plane of rotation, $\bm{n}=\hat{\bm{y}}$. The pure dephasing rate is zero, $J_0=0$. The frequency of rotation of the control field is $\Omega=0.1\Delta$. The coupling to environment $\alpha=0.05$.}
\end{figure}

The Bloch equations Eq.~\eqref{eq:Blochall} with rates given by Eq.~\eqref{eq:ratesA} can be solved to get the qubit density matrix $\hat{\rho}^{U_2}(t)$ in the secondly rotated basis.
In conjunction with the initial condition, the time evolution of $m_y(t)$ is then obtained by ${\rm Tr}[\hat{\Sigma}_y\hat{\rho}(t)]$. 

First, we provide an exact analytical solution by choosing the initial state to be a thermal state $\rho^{(0)}(0) = 1/2+\tanh(W/2T)\sigma_z/2$. Defining $m_0 = \tanh(W/2T)$, the initial condition for the Bloch equation becomes $m_z(0) = m_0\cos\eta$, $m_x(0) = 0$ and $m_y(0) = m_0\sin\eta$. Integrating the Bloch equation with the above initial condition yields
\begin{align}
\label{eq:myanalytic}
m_y(t) & = - m_0 \sin\eta \\
\times & \left(1-2\sin^2\frac{\eta}{2} e^{- \Gamma_{\rm tot} t} - \cos \eta e^{- \Gamma_{\rm tot}t/2} \cos W t\right),\nonumber
\end{align}
where $\Gamma_{\rm tot} = \Gamma_r + \Gamma_e$ and we assumed $J_0=0$. In the long times limit, $t\to\infty$, $m_y(t)$ reaches its stationary state solution 
\begin{equation}
\label{eq:minftyA}
m_y(\infty) = -\frac{\Omega}{W} \tanh\frac{W}{2T}\simeq-\frac{\Omega}{\Delta}\tanh\frac{\Delta}{2T} ,
\end{equation}
regardless of the form of the initial state. The significance of this expression is that the dynamical transverse response of the qubit subject to a rotating magnetic field is a consequence of the geometric phase effect in the sense that the stationary value $m_y(\infty)$ does not depend on the strength of the coupling to environment. 
Therefore, $m_y(\infty)$ is purely geometrical and immune to quantum zero-temperature fluctuations of the environment. 

Next, in order to get the numerical solution of the BR equations \eqref{eq:BRall} we utilize standard integration methods for a system of linear differential equations with time-dependent coefficients.  
Alternatively, we obtain the same results using the BR functions of the QuTiP package~\cite{Johansson2012,Johansson2013} with a proper adjustment to the system Hamiltonian and the interaction term, see  Eqs.~\eqref{eq:HintV} and \eqref{eq:HV}, for time--dependence of the eigenstate basis, 
as presented in Figs.~\ref{fig:3A_my_vs_t_variosgammaTzero} and \ref{fig:3A_my_vs_t_for_temp}.  We verified that the results shown in the plots are identical to numerical integration of the BR equations with the rates given by Eqs.~\eqref{eq:ratesA}. 
In both plots, the initial condition of the density matrix is chosen to be the ground state at $t=0$ when $\bm{b}\| \bm{e}_z$.  We obtain plots consistent with the analytical result, Eq.~\eqref{eq:myanalytic}, for the thermal state of the density matrix at $t=0$. 

In Fig.~\ref{fig:3A_my_vs_t_variosgammaTzero}, we present the time evolution of $m_y(t)$ for several values of the coupling to the environment. From the plot it is clear that the role of the environment is to suppress transient wiggles of $m_y$ and to bring the system to the steady state, defined by Eq.~\eqref{eq:minftyA} with $\tanh(\Delta/2T)\to 1$. 
However, the transverse magnetization is fragile to thermal fluctuations, since these fluctuations create excitation to 
the higher energy state. The result is shown in Fig.~\ref{fig:3A_my_vs_t_for_temp}, where we fix $\alpha$ and plot  $m_y(t)$ for different temperatures $T=\{0, 1/2, 1\}\Delta$. We note that since the dephasing rate, $\Gamma_2 = (\Gamma_r + \Gamma_e)/2$ grows with the temperature, the oscillations decay faster for higher temperatures. Also, at finite temperatures, the spin has nonzero probability to stay in the excited state, the asymptote of $m_y(t\to \infty)$ is reduced in agreement with Eq.~\eqref{eq:minftyA}.

\begin{figure}
\begin{centering}
\includegraphics[width=0.9\columnwidth]{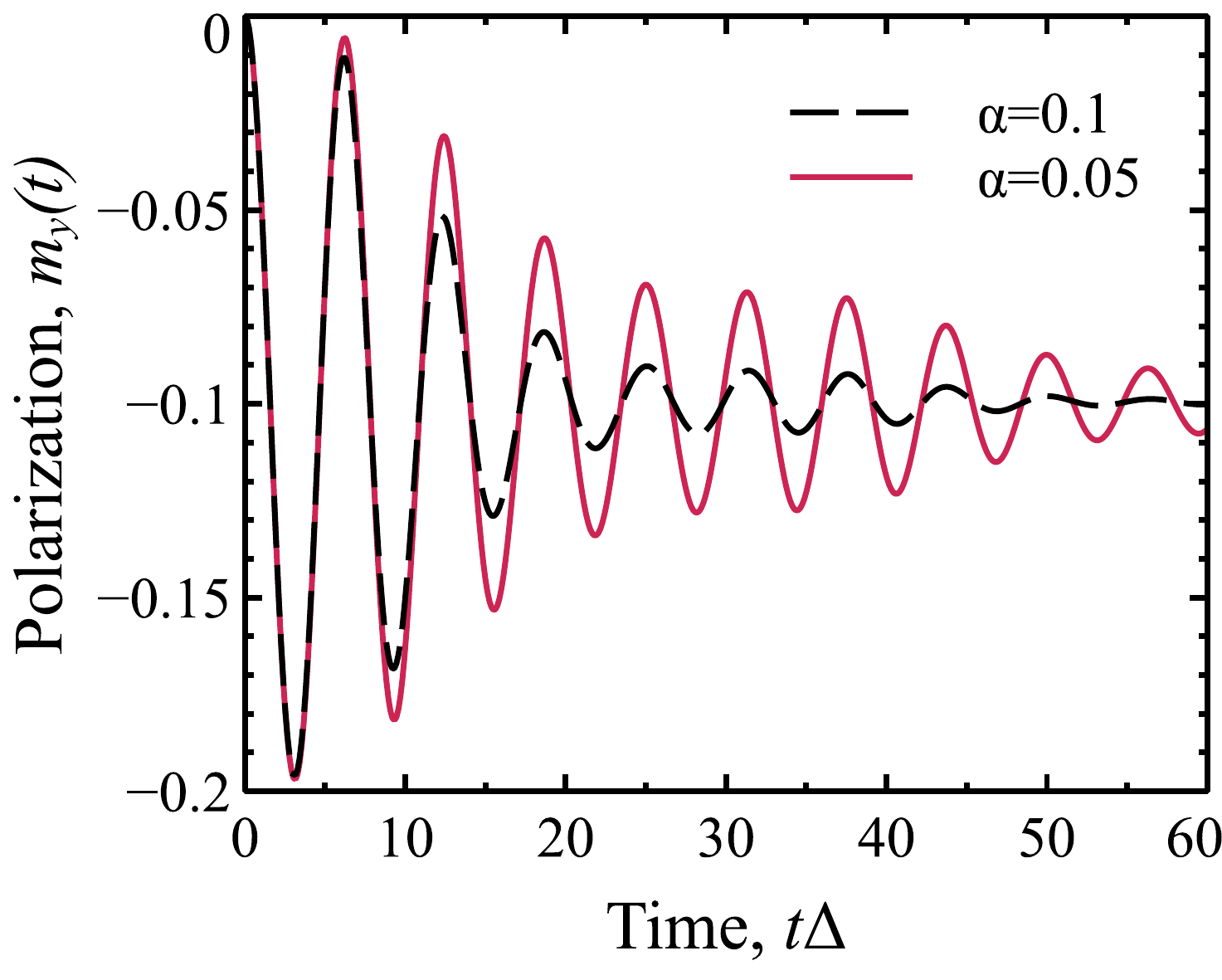}
\end{centering}
\caption{\label{fig:3B_Z_zerotemp_variousgamma}(Color online) Time dependence of the out-of-plane polarization, $m_y(t)$, at zero temperature of environment, for $\alpha=0.05$ (solid line) and $\alpha=0.1$ (dashed line) for fluctuating environment field in the plane of rotation, $\bm{n}=\hat{\bm{z}}$. The pure dephasing rate is zero, $J_0=0$. The frequency of rotation of the control field is $\Omega=0.1\Delta$. The relaxation is reduced for time intervals when $\Omega t\simeq \pi n$.}
\end{figure}

\subsection{Environment field in the rotation plane}

We now consider the qubit interacting with environment field in the plane of rotation.  We take $\bm{n}= \bm{e}_z$ and for $\bm{b}\|\bm{e}_z$ the coupling to the environment results in pure dephasing and is characterized by the low frequency spectral density $J_0$.  As $\bm{b}$ rotates, the effect of environment alternates between pure dephasing and qubit transitions between eigenstates. We obtain this variation in qubit flip and dephasing rates already after applying transformation $\hat{U}_{1}=\exp(i\hat{\sigma}_{y}\theta/2)$ to the interaction Hamiltonian of the qubit and environment, Eq.~\eqref{eq:Hint}. However, for rotating $\bm{b}(t)$ we have to take into account the gauge term $-i\hat{U}_{1}(t)\partial_t{\hat{U}}_{1}^{\dagger}(t)$ in Eq.~\eqref{eq:HU1} by applying the second transformation $\hat{U}_{2}$ to $\hat{H}_{\rm int}$. We obtain
\begin{equation}
\hat{V}(t) {\hat{\sigma}_{z}} \hat{V}^\dagger(t)  =- \hat{\sigma}_{x}\sin\Omega t +  (\hat{\sigma}_y \sin\eta - \hat{\sigma}_z
\cos\eta) \cos\Omega t
\end{equation}
that contains matrix elements for qubit flip processes at any moment of time. 
The corresponding rates in the Bloch--Redfield equations are
\begin{subequations}
\label{eq:ratesZ}
\begin{align}
\Gamma_{r} & = \frac{G(t)}{2}J(W) [N(W)+1], \\
\Gamma_{e} & = \frac{G(t)}{2}J(W)  N(W),\\
\Gamma_{2} & = \frac{\Gamma_{r}+\Gamma_{e}}{2}
+
J_0 \cos^{2}\eta\cos^{2}\Omega t , 
\end{align}
\end{subequations}
where $G(t) \equiv \sin^{2}\eta+\sin^{2}\Omega t\cos^{2}\eta$ and thus the qubit flip rates 
are nonzero as a function of time.

The evolution of the qubit in this case corresponds to precession of a spin in the magnetic field with initial state distinct from its new ground state after the quench.  Namely, its dynamics will correspond to suppression of off-diagonal elements of its density matrix with the rate $\Gamma_2(t)$ and equilibration of the diagonal elements of $\rho$ with rates $\Gamma_{r/e}(t)$.  
We emphasize that in this case all decoherence rates are time-dependent. 

\begin{figure}
\begin{centering}
\includegraphics[width=0.9\columnwidth]{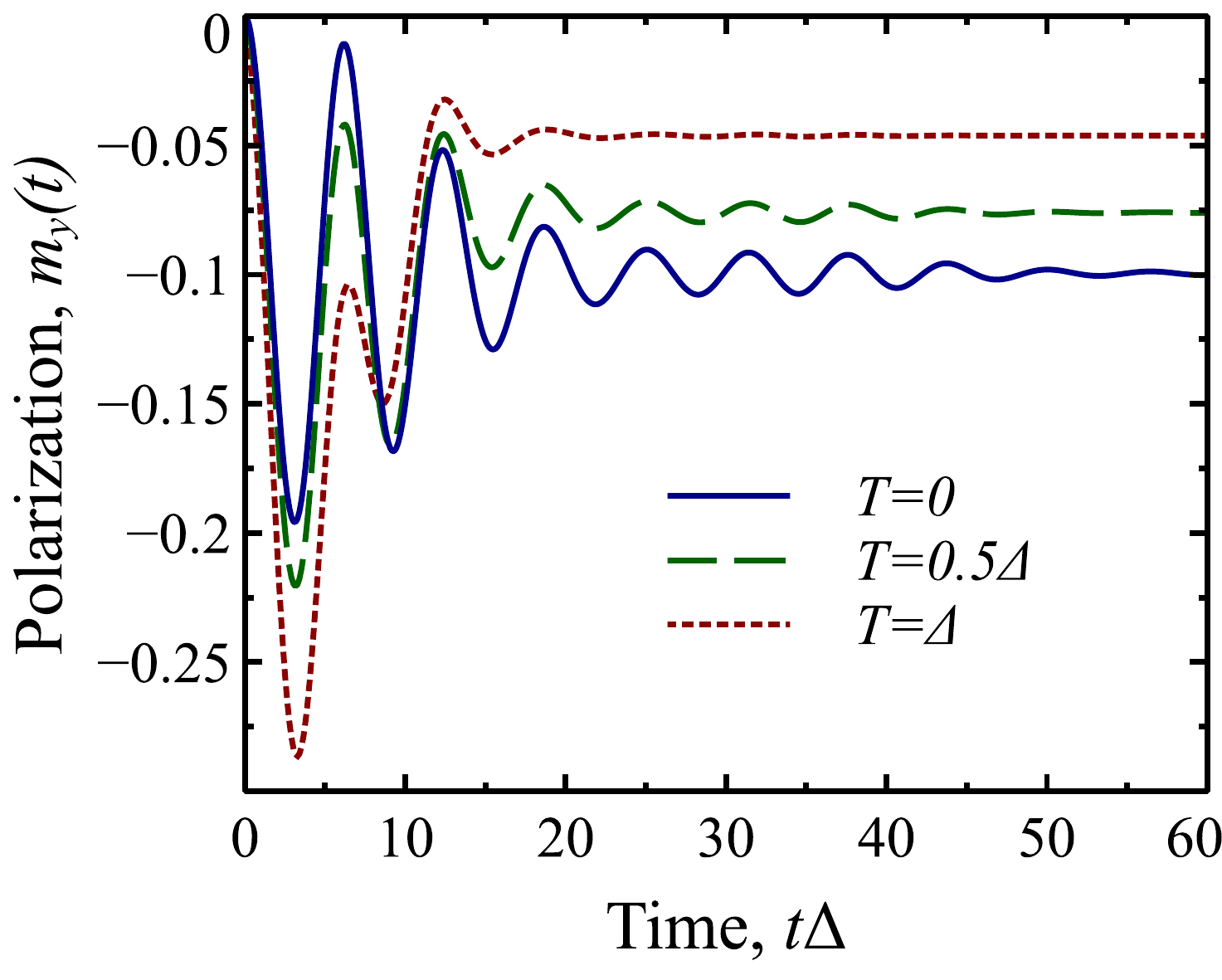}
\end{centering}
\caption{\label{fig:3B_Z_variostemp}(Color online) Time dependence of the out-of-plane polarization, $m_y(t)$, for $\alpha=0.1$ and $T=0$ (solid line), $T=0.5\Delta$ (dashed line) and $T=\Delta$ (dotted line) in case when fluctuating environment field is in the plane of rotation, $\bm{n}=\hat{\bm{z}}$. The pure dephasing rate is zero, $J_0=0$. The frequency of rotation of the control field is $\Omega=0.1\Delta$. 
}
\end{figure}

We calculate time-dependence of $m_y(t)$ by numerically solving the BR equations with the rates given by Eq.~\eqref{eq:ratesZ}.  We present the result of integration in Fig.~\ref{fig:3B_Z_zerotemp_variousgamma} for two different values of $\alpha$ at zero temperature and find clear evidence that the decoherence rates are roughly one half smaller compared to the result of previous subsection for the same value of $\alpha$. Meanwhile, in Fig.~\ref{fig:3B_Z_variostemp} we fix $\alpha$ and plot $m_y(t)$ for different temperatures.  At time longer than the relaxation time $1/\Gamma_2$,  $m_y(t)$
becomes constant with its value $m_y(\infty) = -\Omega/W\tanh(W/2T)$, see Eq.~\eqref{eq:minftyA}

\subsection{Longitudinal coupling to environment}

We also consider a somewhat artificial scenario when the coupling vector $\bm{n}= \sin\Omega t \,\bm{e}_x + \cos\Omega t\, \bm{e}_z$ in Eq.~\eqref{eq:Hint}  rotates together with the external field $\bm{b}(t)$ \footnote{This case may be realized if the interaction of the environment with the qubit is introduced through a fluctuating field along the external field $\bm{b}(t)$, \textit{e.g.} when $\bm{b}(t)$ is realized as two quadratures of microwave pulse driving a qubit and the environment is described by longitudinal quantum fluctuations  of the pulse.}.  For a stationary Hamiltonian this environment does not produce qubit flip processes and results in pure dephasing, when the diagonal elements of the density matrix do not change and only off diagonal elements decrease with time. In case when the direction of the control field rotates with frequency $\Omega$, 
the basis transformation term in Eq.~\eqref{eq:HU1} introduces qubit flip processes for this coupling with the rates in Eqs.~\eqref{eq:BRall} given by
\begin{subequations}
\label{eq:Gammas_LF}
\begin{align}
\Gamma_{r} & = \frac{\sin^2\eta}{2}J(W) [N(W)+1], \\
\Gamma_{e} & = \frac{\sin^2\eta}{2}J(W)  N(W),\\
\Gamma_{2} & = \frac{\Gamma_{r}+\Gamma_{e}}{2}
+
\cos^{2}\eta J_0. 
\end{align}
\end{subequations}
For slow rotation $\Omega\ll \Delta$, we have  $\sin \eta \ll 1$ and qubit flip processes are small.  In this case, dephasing will suppress precession on time scale $\sim 1/J_0$, and further equilibration of the system occurs on a longer time scale $\sim \Delta /\pi\Omega^2$. We describe the evolution of a qubit coupled to high--temperature environment  using a dephasing Lindblad model in Sec.~\ref{sec5}.

\begin{figure}
\begin{centering}
\includegraphics[width=0.9\columnwidth]{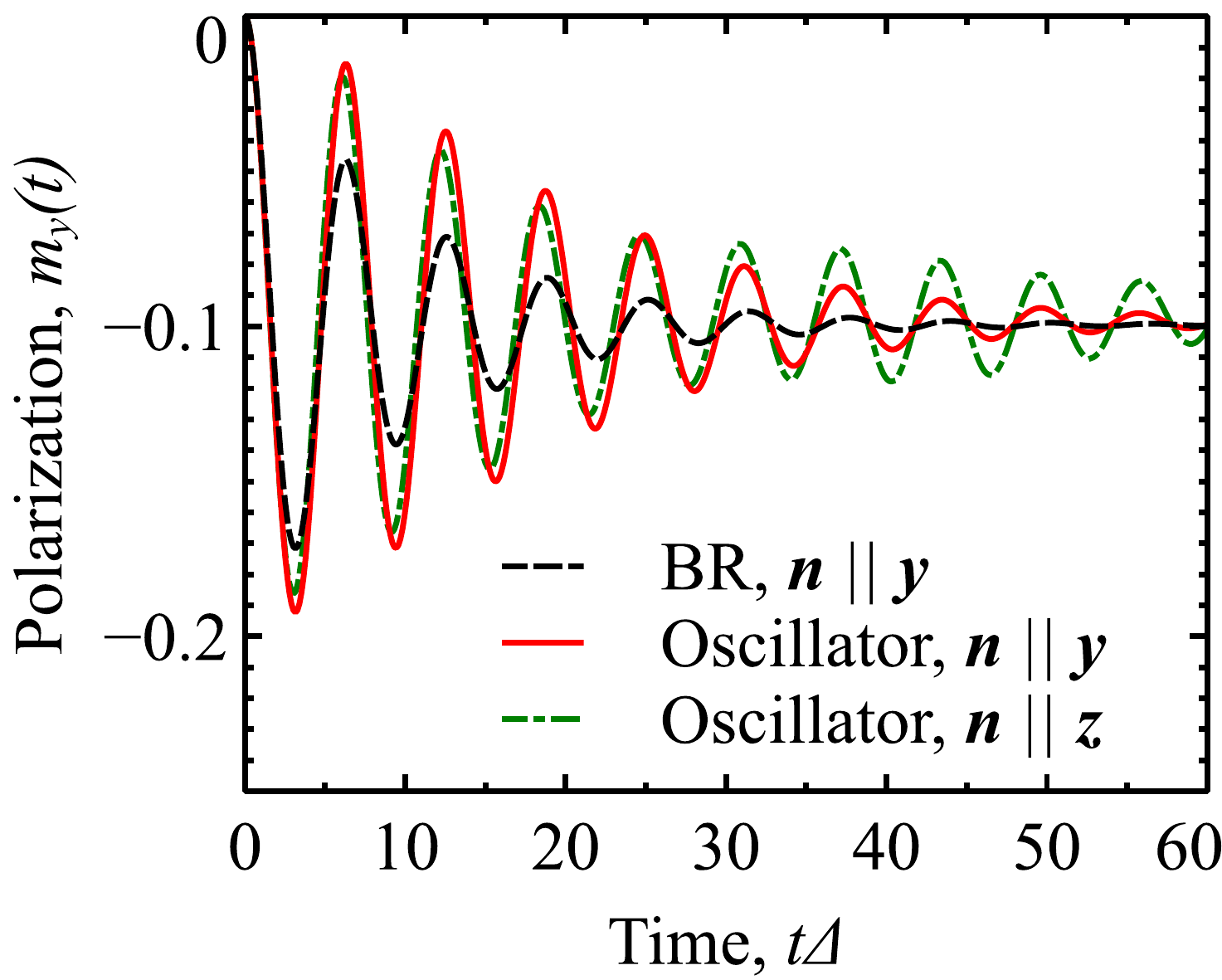}
\end{centering}
\caption{\label{fig:3C_ZEROtemp}(Color online) Time dependence of the out-of-plane polarization, $m_y(t)$, at zero temperature of environment for a qubit coupled to a damped harmonic oscillator with damping rate $\kappa=0.2\Delta$ and coupling constant between the qubit and environment $\lambda=0.1\Delta$. Coupling vector $\bm{n}\parallel\hat{\bm{y}}$ (solid line) and $\bm{n}\parallel\hat{\bm{z}}$ (dash-dotted line).  For comparison, the solution for the Bloch--Redfield equation is presented (dashed line) with $\alpha=0.03$ and $J_0=0$.  The rotation angular velocity is $\Omega=0.1\Delta$.
}
\end{figure}

\subsection{Coupling to a strongly damped Quantum oscillator}

In this subsection we consider the interaction of a qubit with a single damped quantum harmonic oscillator.  This model can be used to describe environment with a sharp spectral function $J(\epsilon)$.  The interaction part of the Hamiltonian 
is similar to Eq.~\eqref{eq:Hint}:
\begin{equation}
\hat{H}_{\rm int}=\frac{\lambda}{2}(\hat{a}+\hat{a}^\dagger)\bm{n} \cdot \bm{\sigma}
\end{equation}
and the single-mode Hamiltonian of the oscillator is $\hat{H}_{\rm o/c}=\omega_0(\hat{a}^\dagger \hat{a}+1/2)$.  We describe dissipation of the oscillator using the Lindblad relaxation operators for the full density matrix 
$\bar\rho(t)$ of the qubit and the oscillator system:
\begin{equation}
\label{eq:hosc_me}
\dot {\bar \rho} (t) = -i[\hat{H}(t), \bar \rho]
-\kappa \left(
\hat{a}^\dagger \hat{a} \bar \rho+  \bar \rho \hat{a}^\dagger \hat{a}- 2 \hat{a}\bar \rho \hat{a}^\dagger
\right) 
\end{equation}
This equation is a standard Lindblad master equation with time dependent Hamiltonian.   
The difference with the previous calculations of this Section is that we keep a full quantum mechanical treatment of the qubit interaction with the oscillator and perform all transformations of the qubit basis for the full Hamiltonian of the qubit and the oscillator. At the same time, we assume that the Lindblad superoperator for the relaxation of the harmonic oscillator, represented by the last term in Eq.~\eqref{eq:hosc_me}, is not affected by these transformations.

\begin{figure}
\begin{centering}
\includegraphics[width=0.9\columnwidth]{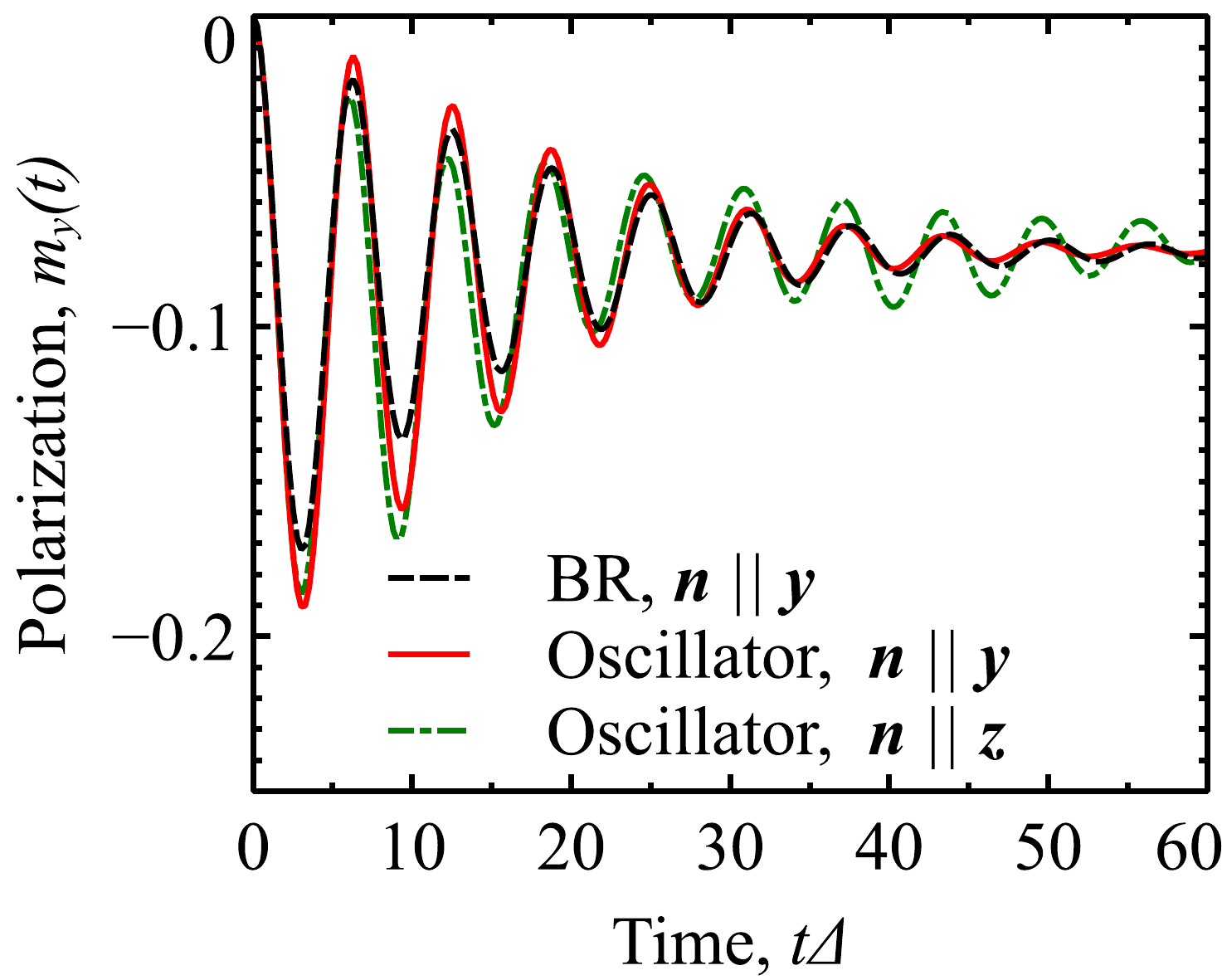}
\end{centering}
\caption{\label{fig:3C_temp0p5}(Color online) Time dependence of the out-of-plane polarization, $m_y(t)$, at environment temperature $T=0.5\Delta$ for a qubit coupled to a damped harmonic oscillator with damping rate $\kappa=0.2\Delta$ and coupling constant between the qubit and environment $\lambda=0.1\Delta$. Coupling vector $\bm{n}\parallel\hat{\bm{y}}$ (solid line) and $\bm{n}\parallel\hat{\bm{z}}$ (dash-dotted line).  For comparison, the solution for the Bloch--Redfield equation is presented (dashed line) with $\alpha=0.03$ and $J_0=0$.  The rotation angular velocity is $\Omega=0.1\Delta$.
}
\end{figure}

We evaluate the qubit projection perpendicular to the rotation plane of the control field as a function of time.  
Fig.~\ref{fig:3C_ZEROtemp} shows the comparison between calculation of Bloch-Redfield equations and damped quantum oscillator with different coupling directions at zero temperature. All three curves saturate at universal value  $m_y(\infty) = -\Omega/W$. It is worth pointing out that the $\bm{n}\parallel\hat{\bm{z}}$ coupling results in time-dependent transition rates that are at minimum when $\bm{b}\parallel \bm{n}$ and at maximum when $\bm{b}\bot \bm{n}$, as one can conclude from the amplitude of oscillations of $m_y(t)$ for $\bm{n}\|\bm{e}_z$.  Effectively, the overall relaxation is slower than that of the case $\bm{n}\parallel\hat{\bm{y}}$ and the amplitude of oscillating $m_y$ at $t\Omega= n\pi $ decays insignificantly.  The calculations at finite temperature $T = 0.5\Delta$ are plotted in Fig.~\ref{fig:3C_temp0p5} and in all cases  $m_y(\infty)$ is consistent with Eq.~\eqref{eq:minftyA}.

\section{Landau--Zener Transition}
\label{sec4}
In this section we consider the Landau--Zener transition in a qubit coupled to its environment.  The external field in the qubit Hamiltonian \eqref{eq:H0} has the following form $\bm{b}(t)=\{\Delta,0,v t\}$, where $\Delta$ is the minimal level separation and $v$ characterizes the rate at which the Hamiltonian changes.  
For the Landau--Zener problem, the qubit is initially in the ground state $|g\rangle$ with the density matrix $\hat{\rho}(t\to-\infty)=|g\rangle\langle g|$.   The task is to find the probability of the system to be in the excited state $|e\rangle$ which is given by $P_\infty=\lim_{t\to +\infty} \langle e| \hat{\rho}(t)|e\rangle$.

\begin{figure*}
  \begin{centering} 
  \includegraphics[width=14cm]{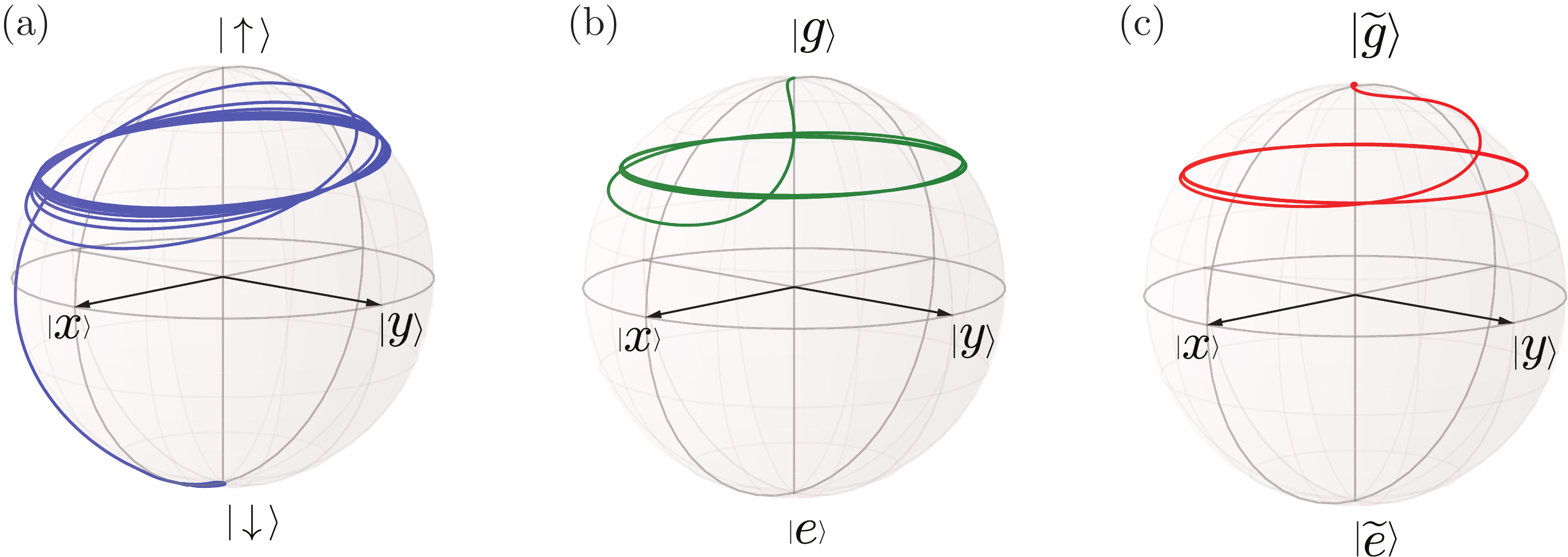}
  \caption{(Color online) Representation of a qubit state during the Landau--Zener process on the Bloch sphere in (a) diabatic basis of states $|\uparrow\rangle$ and $\downarrow\rangle$ along fixed $\hat{\bm{z}}$ axis; (b) adiabatic basis of the ground, $|g\rangle$, and excited, $|e\rangle$, states; (c) in the ``improved'' eigenstate basis, $|g\rangle$ and $|e\rangle$, obtained from the diabetic basis by $U_2$ transformation.  In diabatic basis the trajectory of qubit state moves across the sphere from state $|\downarrow\rangle$ (ground state at $t\to-\infty$) towards
$|\uparrow\rangle$ and slowly approaches the circle of constant precession at $t\to\infty$.  The trajectory in the adiabatic basis and eigenstate basis shows a simpler trajectory and fast switch to the constant precession circle.  Level-crossing speed $v=0.5\Delta^2$ and no coupling to environment.\label{fig:3spheres}}
  \end{centering}
\end{figure*}

Effects of the environment on qubit's dynamics can be separated into  pure dephasing of the qubit state during the LZ process and inelastic qubit flips.  When we consider a qubit coupled to its environment that causes qubit flip processes, we have to be careful with the formulation of the LZ problem.   Indeed, the LZ process is formally infinitely long and the qubit flip processes accompanied by the energy exchange will result in equilibration of the qubit system with its environment.  In particular, for the zero temperature environment, the qubit will relax to the ground state even if it was temporarily excited during the LZ process.  For environment at finite temperature, the qubit state will tend to thermal state $\hat{\rho}(t)=\mathrm{diag}\{\rho_{00},\rho_{11}\}$ with $\rho_{11}/\rho_{00}=\exp(-E(t)/T)$.  But as formally $E(t)\to \infty$ for long times $t$, the qubit will relax to the ground state and we find $P_{\infty}=0$. 

Previous considerations, see \textit{e.g.}~\cite{Wubs2006}, 
 predicted $P_{\infty}\to 0$ for the Ohmic environment with large high-frequency cutoff in the environment modes.\footnote{Notice that the projection of the coupling vector between a qubit and the environment does not change in time in Ref.~\cite{Wubs2006} as required for the proper treatment of environment--qubit interaction.}   
But in this case the problem looses its meaning since the LZ transition is shadowed by trivial relaxation of a quantum system to its ground state by releasing its energy to the environment. One can reformulate the problem in terms of finite time LZ process, which may be experimentally relevant situation in some cases.  Alternatively, one can assume that the environment spectral function has a relatively low cutoff at high frequencies $E_c\sim \Delta$ and the relaxation is absent after time $t\gtrsim E_c/v$.  Here, we consider a special orientation of the coupling vector with environment when $\bm{n}\| \bm{e}_z$,where $\bm{e}_z$ is defined by $\hat{\bm{b}}(t\to\pm \infty)\| {\bm{e}}_z$.  In this situation, the relaxation processes becomes weak at long times  $|t|\gg \Delta / v$.  This type of coupling is expected to be dominant in qubits with relatively long energy relaxation times, but with short dephasing time due to dominant coupling with the fluctuating field parallel to the qubit field along $\bm{e}_z$.  

We utilize the Bloch--Redfield approach to the problem of Landau--Zener transitions in the presence of environment with $\bm{n}=\bm{e}_z$.  In 
principle, we need to write the BR equations in the basis where the transformed qubit Hamiltonian  is diagonal after an infinite series of basis transformations given by $\hat{U}_n$, which can be an infinite series. However, under the condition $v\lesssim \Delta ^2$, the series of basis transformations can be limited by $\hat{U}_2(t) \hat{U}_1(t)$.
 
\begin{figure}
  \begin{centering}
  \includegraphics[width=0.9\columnwidth]{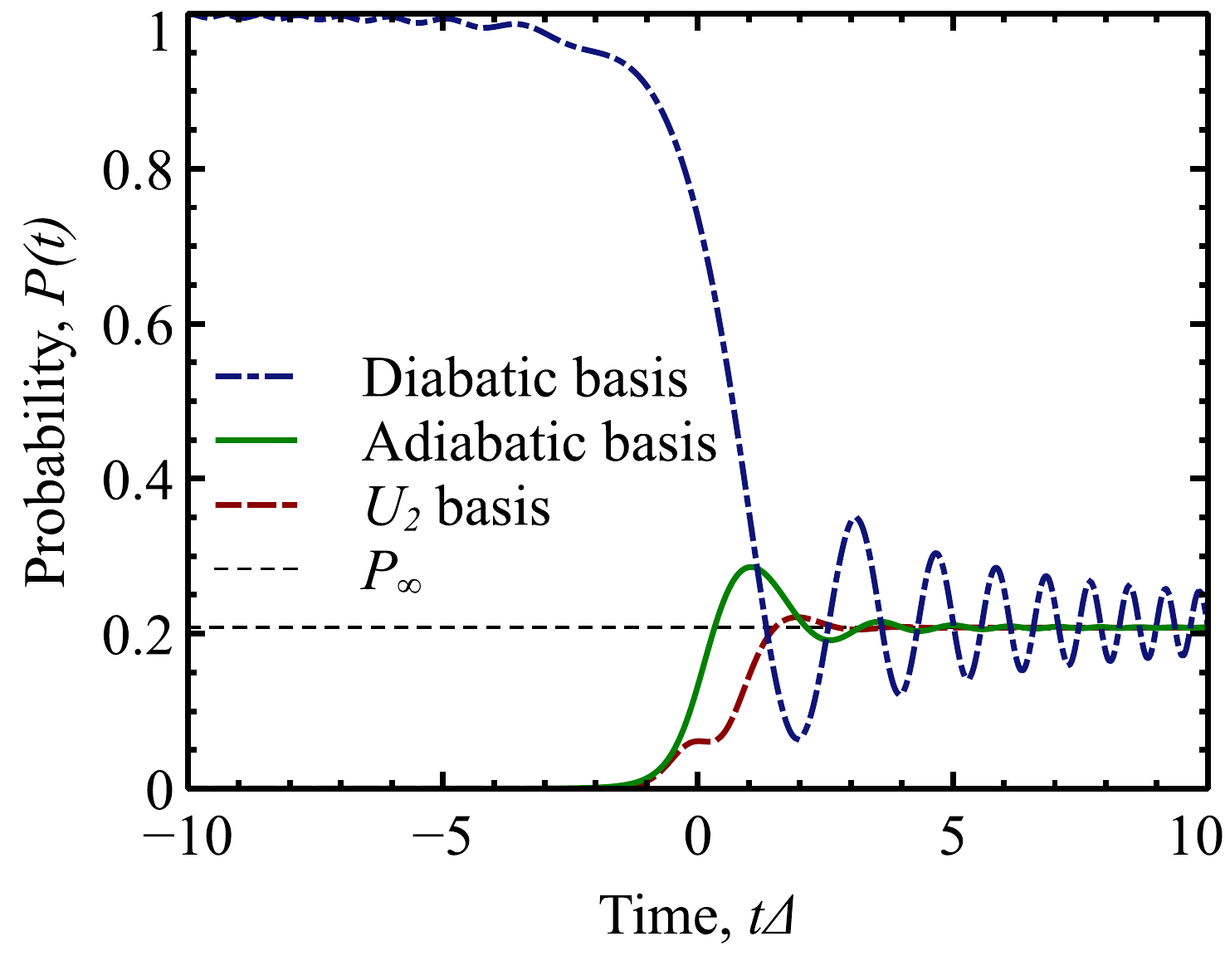}
  \caption{\label{fig:LZidealPz}(Color online) projection of a qubit state during the Landau--Zener process on the Bloch sphere on $|\downarrow\rangle$ state in the diabatic basis (dash-dotted line) and on the ground state in the adiabatic basis (solid line) and the ``improved'' eigenstate basis (dashed line).  In the diabatic basis the projection of the qubit state shows long oscillations with amplitude decreasing as a power law in time, while the eigenstate projections quickly reach the asymptotic value.  Level-crossing speed $v=0.5\Delta^2$ and no coupling to environment.}
  \end{centering}
\end{figure}

The first  transformation changes the representation from diabatic basis of states $|\uparrow\rangle$ and $|\downarrow\rangle$ along $\bm{e}_z$ to the  adiabatic basis of the ground, $|g\rangle$, and excited, $|e\rangle$, states, where the Hamiltonian is diagonal.  
The first transformation matrix $\hat{U}_1(t)$ has the same form as in Eq.~\eqref{eq:u1} except the rotation angle $\theta(t)$, which is now defined as
\begin{equation}
\label{eq:costheta}
\cos\theta(t)=-\frac{vt}{\sqrt{v^2t^2+\Delta^2}}.
\end{equation}
The transformed Hamiltonian in the adiabatic basis has the form\cite{Ao1991,Shevchenko2004,Vitanov1999,Grifoni1998}
\begin{equation}
\label{eq:H0U1LZ}
\hat{H}_0^{U_1}=-\frac{E(t)}{2}\hat{\sigma}_z-\frac{v\Delta}{2E^2(t)}\hat{\sigma}_y,\quad 
E(t)=\sqrt{v^2t^2+\Delta^2}.
\end{equation}

The second transformation is chosen to diagonalize matrix $\hat{H}_0^{U_1}$ and has the form 
\begin{equation}
\label{eq:U2LZ}
\hat{U}_2(t)=\exp\left(-\frac{i\eta}{2}\hat{\sigma}_x\right), \quad \tan \eta(t)=\frac{v\Delta}{E^3(t)}.
\end{equation} 
The Hamiltonian in this ``improved eigenstate'' basis has the form
\begin{subequations}
\label{eq:H0U2LZ}
\begin{align}
\hat{H}_0^{U_2} & =-\frac{W(t)}{2}\hat{\sigma}_z-\frac{\dot\eta}{2}\hat{\sigma}_x,
\label{eq:H0U2LZ_a}
\\
W(t)& =\sqrt{E^2(t)+\frac{v^2\Delta^2}{E^4(t)}}, \quad \dot\eta = \frac{3v^3\Delta t}{E^3(t)W^2(t)}
\label{eq:H0U2LZ_b}.
\end{align}
\end{subequations}

Without dissipation, the LZ problem is equivalent in all three representations, with a properly written Hamiltonian, i.e., Eq.~\eqref{eq:H0} for the diabatic basis,
Eq.~\eqref{eq:H0U1LZ} for the adiabatic basis, and Eq.~\eqref{eq:H0U2LZ_a} for ``improved eigenstate'' basis.  In all representations, the qubit follows the appropriate instantaneous control field $\tilde{\bm{b}}(t)$, but since this field is time-dependent, the qubit deviates from the instantaneous direction of $\tilde{\bm{b}}(t)$ and acquires an additional precession around the control field.  When the original field eventually reaches its final direction, $\bm{b}\|\bm{e}_z$ at $t\gg \Delta/v$, the direction of the control field becomes time independent and the qubit simply precesses around $\bm{e}_z$ with a non-zero projection of its state on the excited state, given by the known expression\cite{Landau1932,Zener1932,Majorana1932}
\begin{equation}
\label{eq:LZideal}
P^{LZ}_\infty=\exp\left(-\frac{\pi \Delta^2}{2v}\right).
\end{equation}
Note that in Fig.~\ref{fig:3spheres} this precession remains in all three considered representations, but the overall trajectories are smoother in the transformed representations.  As we look at the projection of the qubit state on the ``excited state'' $P(t)=\langle e| \hat{\rho}(t) | e\rangle$ in the appropriate basis, see Fig.~\ref{fig:LZidealPz}, the oscillations decrease faster in the transformed representations, because the control field $\bm{b}(t)$ aligns faster with its final direction.  We also note that since the control field remains aligned with its initial direction longer in transformed basis, the numerical computation can run over shorter time intervals thus making computation faster and more accurate.

Next, we take into account interaction with the environment within the Bloch--Redfield approach. The coupling to the environment is modified in the diagonal basis of the Hamiltonian, see Eq.~\eqref{eq:HintV} and Ref.~\cite{Grifoni1998}. Under the Markovian approximation and to the second order in the coupling to environment, we obtain the corresponding  BR equations in the form\begin{subequations}\label{eq:BRLZ}
\begin{align}
\dot \rho_{00} & = i\frac{\dot\eta}{2} (\rho_{01}-\rho_{10})-\Gamma_e\rho_{00}+\Gamma_r\rho_{11}, \\   
\dot \rho_{11} & = -i\frac{\dot\eta}{2} (\rho_{01}-\rho_{10})+\Gamma_e\rho_{00}-\Gamma_r\rho_{11}, \\
\dot \rho_{01} & = -(\Gamma_2 +iW(t)) \rho_{01} + i\frac{\dot\eta}{2} (\rho_{00}-\rho_{11}),\\
\dot \rho_{10} & = -(\Gamma_2 -iW(t)) \rho_{10} - i\frac{\dot\eta}{2} (\rho_{00}-\rho_{11}),
\end{align}
\end{subequations}
where $W(t)$ and $\dot\eta$ are given by Eq.~\eqref{eq:H0U2LZ_b}. 
The rates for the above equations are
\begin{subequations}
\label{eq:BRratesLZ}
\begin{align}
\Gamma_{r} & = \frac{G_{LZ}(t)}{2}J(W(t)) [N(W(t))+1], 
\label{eq:GammarLZ} 
\\
\Gamma_{e} & = \frac{G_{LZ}(t)}{2}J(W(t))  N(W(t)),\\
\Gamma_{2} & = \frac{\Gamma_{r}+\Gamma_{e}}{2}
+
J_0 \cos^{2}\eta\cos^{2}\theta(t) , 
\end{align}
\end{subequations}
where $G_{LZ}(t)=\sin^{2} \eta+\sin^{2}\theta(t)\cos^{2} \eta$  is a function of time--dependent basis rotation angles $\theta(t)$ and $\eta(t)$ defined by Eqs.~\eqref{eq:costheta} and \eqref{eq:U2LZ}. 
We note that the above equations for BR rates are given by truncation of transformation series of interaction Hamiltonian, Eq.~\eqref{eq:HintV}, up to the second order,  $\hat{V} = \hat{U}_2(t)\hat{U}_1(t)$.
Therefore, the rates are defined  within ${\cal O}(\eta^2) \lesssim {\cal O}(v^2/\Delta^4)$ accuracy.  The unitary evolution described by either $\hat{H}_0^{U_1}$ or  $\hat{H}_0^{U_2}$ has no approximations and is valid for arbitrary values of $v$.  We  emphasize that once the basis transformation gives rise to non-zero decoherence rates, the qualitative results are similar regardless of our choice of the BR rates in the basis  obtained after either $\hat{U}_1$ or $\hat{U}_2\hat{U}_1$ transformations. The rates in the $\hat{U}_1$ basis are given by Eq.~\eqref{eq:BRratesLZ} with $\eta=0$. We now discuss solution of Eq.~\eqref{eq:BRLZ}. 
   
\begin{figure}
\begin{centering}
\includegraphics[width=0.9\columnwidth]{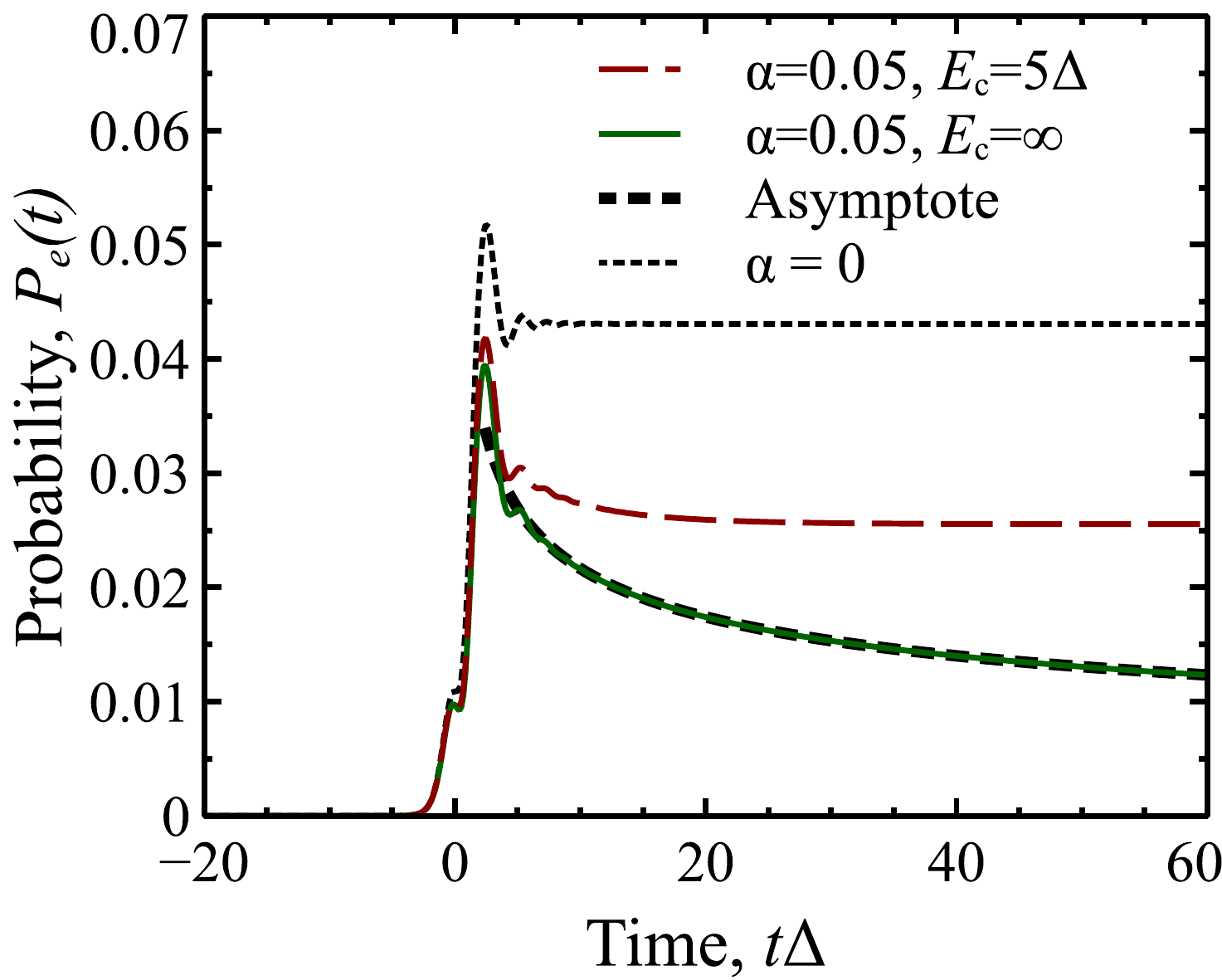}
\end{centering}
\caption{\label{fig:LZ_T0}(Color online) The probability of occupation of the excited state in the Landau--Zener transition in the $U_2$ basis. The temperature of environment is zero, $T=0$, the level velocity is $v=0.5\Delta^2$.  We assume that the dephasing is absent, $J_0=0$.  The asymptotic curve for $E_c=\infty$ is given by Eq.~\eqref{eq:rho11LZT0b} with a proper choice of integration constant $C$. 
}
\end{figure}

\subsection{Zero temperature dissipative environment}

We first consider the zero temperature environment and set $J_0 = 0$ to focus solely on qubit flips rather than dephasing. We numerically integrate the BR equation \eqref{eq:BRLZ} and plot 
the probability of the system to be in the excited state  $P_e(t)=\langle e|\rho (t)| e\rangle $ as a function of time in Fig.~\ref{fig:LZ_T0} for $\alpha=0.05$.  For numerical integration, we used both direct integration of linear differential equations \eqref{eq:BRLZ} and the QuTiP's package for numerical solution of the Bloch--Redfield equations \cite{Johansson2012,Johansson2013}, obtaining identical results.
As the qubit levels go over the avoided crossing, the probability of the qubit to be in the excited state increases, roughly following the same function of time as $P_e(t)$ for an isolated qubit, $\alpha=0$. As the levels further depart from each other, the relaxation of the qubit from the excited state becomes the dominant process in the qubit dynamics, and  $P_e(t)$ monotonically decreases and becomes constant once the level separation $\sim vt $ exceeds the ultraviolet cutoff $E_c$, or $t\gtrsim E_c/v$ and the  qubit is effectively decoupled from the environment.
In Fig.~\ref{fig:LZ_T0} we compare the behavior of $P_e(t)$ for different values of 
$E_c$. For finite ultraviolet cutoff $E_c=5\Delta$, the probability $P_e(t)$ saturates for $t\Delta\gtrsim  10$.
For $E_c\to\infty$, the probability $P_e(t)$ slowly decreases for all $t>\Delta/v$.

To evaluate this suppression, we can utilize Eqs.~\eqref{eq:BRLZ} in the asymptotic regime for $t\gg v/\Delta$, when $\Gamma_r(t)\gg \dot\eta(t)$. We write 
\begin{subequations}
\label{eq:rho11LZT0}
\begin{align}\label{eq:rho11LZT0a}
\frac{dP_e(t)}{dt} & =- \frac{\Delta^2}{2v^2t^2}J(vt) P_e(t),\\
\label{eq:rho11LZT0b}  
P_e(t) & = C\exp\left(- \frac{\pi\alpha\Delta^2}{v}\ln \frac{vt}{\Delta}\right) \propto 
t^{-\pi\alpha\Delta^2/v}.
\end{align}  
\end{subequations}
where we used the relaxation rate $\Gamma_r$ from Eq.~\eqref{eq:GammarLZ}.  The latter equation demonstrates that even for environment with $\bm{n}\|\bm{e}_z$, the relaxation on long times scales is important.  Formally, the power--law dependence of $P_e(t)$ on time originates from the slow converging integral $\int dt J(vt)/v^2t^2 \sim \ln vt/\Delta$ due to linearly increasing environment spectral function $J(\epsilon)$ with energy.   With a proper choice of integration constant C, we obtain a good agreement between computed $P_e(t)$ in Fig.~\ref{fig:LZ_T0} and asymptote, defined by Eq.~\eqref{eq:rho11LZT0b}.

This power--law dependence stops and reaches a fixed value $P_\infty$ when the qubit level separation exceeds the environment ultra-violet cutoff at times $t\gtrsim E_c/v$.
We evaluate the long time asymptotic value of $P_e(t\gg E_c/v)=P_\infty$ by taking into account the high  energy cutoff in the environment spectral function, Eq.~\eqref{eq:J}.  We obtain
\begin{equation}
\label{eq:PinftyT0}
P_\infty(T=0)= C \Pi,\quad \Pi=\exp\left\{ -\frac{\pi \alpha \Delta^2}{v}
\ln\frac{2 E_{c}}{e^\gamma \Delta}
\right\},
\end{equation}
where $\gamma \simeq 0.577$ is the Euler's constant, the integration constant $C\sim P_\infty^{LZ}$ and factor $\Pi $ describes suppression of the excited state due to slow relaxation while qubit level separation increases from its minimum $\Delta$ to values above the cutoff energy $E_c$, see Appendix A for the derivation of Eq.~\eqref{eq:PinftyT0}.

Equations \eqref{eq:rho11LZT0} are valid for $\alpha\ll 1$.  For larger values of $\alpha$, one has to take into account the renormalization of qubit Hamiltonian when the off-diagonal matrix element in the original Hamiltonian $\Delta_r$ is given by  the following self-consistent relation\cite{Leggett1987}
\begin{equation}
\Delta_r = \Delta \exp{\left(-\frac{1}{2}\int_0^{\infty}\frac{J(\omega)}{\omega^2 - \Delta_r^2}d\omega\right)}
\end{equation}
with solution $\Delta_r=\Delta(\Delta/E_c)^{\alpha/(1-\alpha)}$. Hence the relaxation rate is\cite{Orth2010}  
\begin{equation}
\Gamma_r(E) = \frac{\pi \Delta_r}{2\Gamma(2\alpha)}\left(\frac{E}{\Delta_r}\right)^{2\alpha-1}
\end{equation}
where $\Gamma(x)$ is the gamma--function.  The integration over time with $E(t)\simeq vt$ gives\cite{Orth2010}
\begin{equation}
\rho_{11}(t)  = C' \exp\left(
-\frac{\pi\Delta_r^2}{4\alpha\Gamma(2\alpha)v}
\frac{(vt)^{2\alpha}}{\Delta^{2\alpha}_r}
\right).
\label{eq:rho11orth}
\end{equation}
Notice  that in the limit $\alpha\ll 1$, $\Delta_r=\Delta$, the relaxation rate $\Gamma_r$ reduces to $\Gamma_r(E)=\pi\alpha \Delta^2/E$ in agreement with the relaxation rate in Eq.~\eqref{eq:rho11LZT0a}.  Similarly, Eq.~\eqref{eq:rho11orth} becomes Eq.~\eqref{eq:rho11LZT0b}

\begin{figure}
\begin{centering}
\includegraphics[width=0.9\columnwidth]{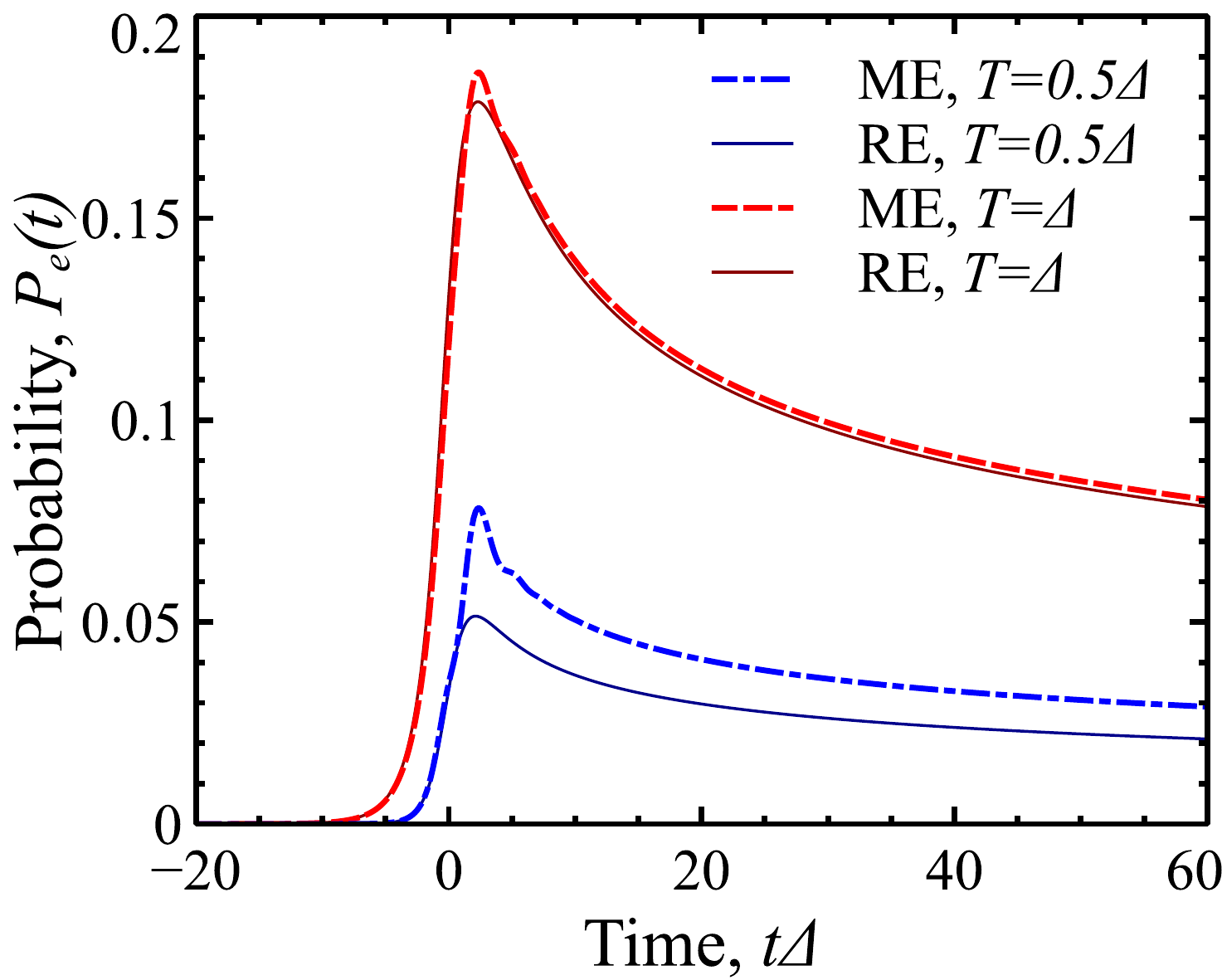}
\end{centering}
\caption{\label{fig:LZ_TempRateEq}(Color online) The probability of occupation of the excited state, $P_e(t)$ in the Landau--Zener transition in the $U_2$ basis at finite temperature of environment for $v=0.5\Delta^2$, $\alpha=0.05$ and  $J_0=0$.  The solid lines represent solutions of rate equations~\eqref{eq:dmdt} that show good agreement with the BR equations at higher temperatures.}
\end{figure}

\subsection{Finite temperatures}

At finite temperatures, the excitation and relaxation rates may exceed $\dot \eta$ terms for strong enough coupling of the qubit to its environment and slow drive $v$.  In this case, we disregard $\dot\eta$ terms in Eq.~\eqref{eq:BRLZ} and the diagonal elements of the density matrix satisfy the rate equations. Since the rate equations preserve the trace of the density matrix, $P_g(t)+P_e(t)=1$, with $P_g(t)=\langle g|\rho(t)|g\rangle$, we introduce $m(t)=P_g(t)-P_e(t)$ and obtain the differential equation for $m(t)$:
\begin{equation}
\begin{split}
\label{eq:dmdt}
\frac{1}{\Gamma_0}\frac{dm}{dt} & = 1 -m \coth\frac{W(t)}{2T}  ,\quad  
\Gamma_0 = \pi\alpha W(t)G_{LZ}(t),\\
G_{LZ}(\tau) & = \frac{\Delta^2(v^2+(v^2\tau^2+\Delta^2)^2)}{v^2\Delta^2+(v^2\tau^2+\Delta^2)^3}.
\end{split}
\end{equation}
The initial condition is $m(t_i)=1$ for $t_i=-\infty$.  While we can write a formal solution to Eq.~\eqref{eq:dmdt}, the solution is not well defined due to logarithmic divergence of $\int_{t_i}\Gamma_0(t)dt$ for the spectral function $J(\varepsilon)$ without a cutoff.   We present the result of numerical solution of Eqs.~\eqref{eq:BRLZ} and the rate equations in Fig.~\ref{fig:LZ_TempRateEq}.  We notice that for higher temperatures, these two solutions are indistinguishable because the thermal effects dominate only in short time scales $|vt|<T$ such that the time window is long enough for the qubit to be thermalized and its off-diagonal elements of density matrix vanish.

Integrating Eq.~\eqref{eq:dmdt} over $t$ yields the following solution of $P_{\infty} = 1/2 - m(\infty)/2$:
\begin{equation}
\label{eq:finiteT}
P_{\infty}=\int_{-\infty}^{\infty}\Gamma_{e}(t)e^{-\int_{t}^{\infty}\Gamma_{0}(t')\coth\frac{W(t')}{2T}dt'}dt.
\end{equation}
The integral over time $t$ is understood as thermal activation processes with rate $\Gamma_e(t)$ and integral in the exponent can be considered as contribution of relaxation processes after thermalization.
For weak coupling $\alpha\ll 1$ and not very high temperatures $\alpha T\ll v/\Delta$, the integral in the exponential is a slow function of $t$. Therefore, we can replace the lower bound of the integration by $t=0$. We obtain $P_{\infty}$  in the limit of low temperatures $T\ll \Delta$
\begin{equation}
\label{eq:finiteTlow}
P_{\infty}\simeq
\frac{2\pi \alpha\Delta^2}{v}\sqrt{\frac{\pi T}{2\Delta}} 
e^{-\Delta/T} \Pi ,
\end{equation}
and in the limit of higher temperatures  $T\gg \Delta$
\begin{equation}
\label{eq:finiteThigh}
P_{\infty}\simeq\frac{2\pi^2\alpha T \Delta}{v} \Pi,
\end{equation}
where  $\Pi$ is defined by Eq.~\eqref{eq:PinftyT0}. 
The details of the derivation of the above equations are presented in Appendix \ref{App:A}. 
We remind that Eqs.~\eqref{eq:finiteT} -- \eqref{eq:finiteThigh} are valid when the rate equations~\eqref{eq:dmdt} are a good approximation to the BR equations~\eqref{eq:BRLZ}.  In this case, the transition of the system to the excited state is a consequence of incoherent excitation by environment of the qubit, and is not the coherent phenomenon that leads to the excitation in the Landau--Zener transition of an isolated quantum system.  However, the excitation processes only happen when the adiabatic eigenstates of the qubit have a non-zero matrix elements with the coupling to environment, the latter happens when the ``control field'' $\hat{\bm{b}}$ is not parallel to the environment field which happens during time  $\Delta /v$, when the excitation rate can be estimated as $\pi\alpha T$, resulting in the excitation probability $\propto \alpha T \Delta/v$, \emph{cf.} to Eq.~\eqref{eq:finiteThigh}. As the level separation $E(t)$ exceeds temperature, only relaxation process remains that causes transitions to the ground state.  The effect of this relaxation is represented by the exponential factor in Eqs.~\eqref{eq:finiteTlow} and \eqref{eq:finiteThigh}, \emph{cf.} to Eq.~\eqref{eq:rho11LZT0b}.  

\begin{figure}
\begin{centering}
\includegraphics[width=0.9\columnwidth]{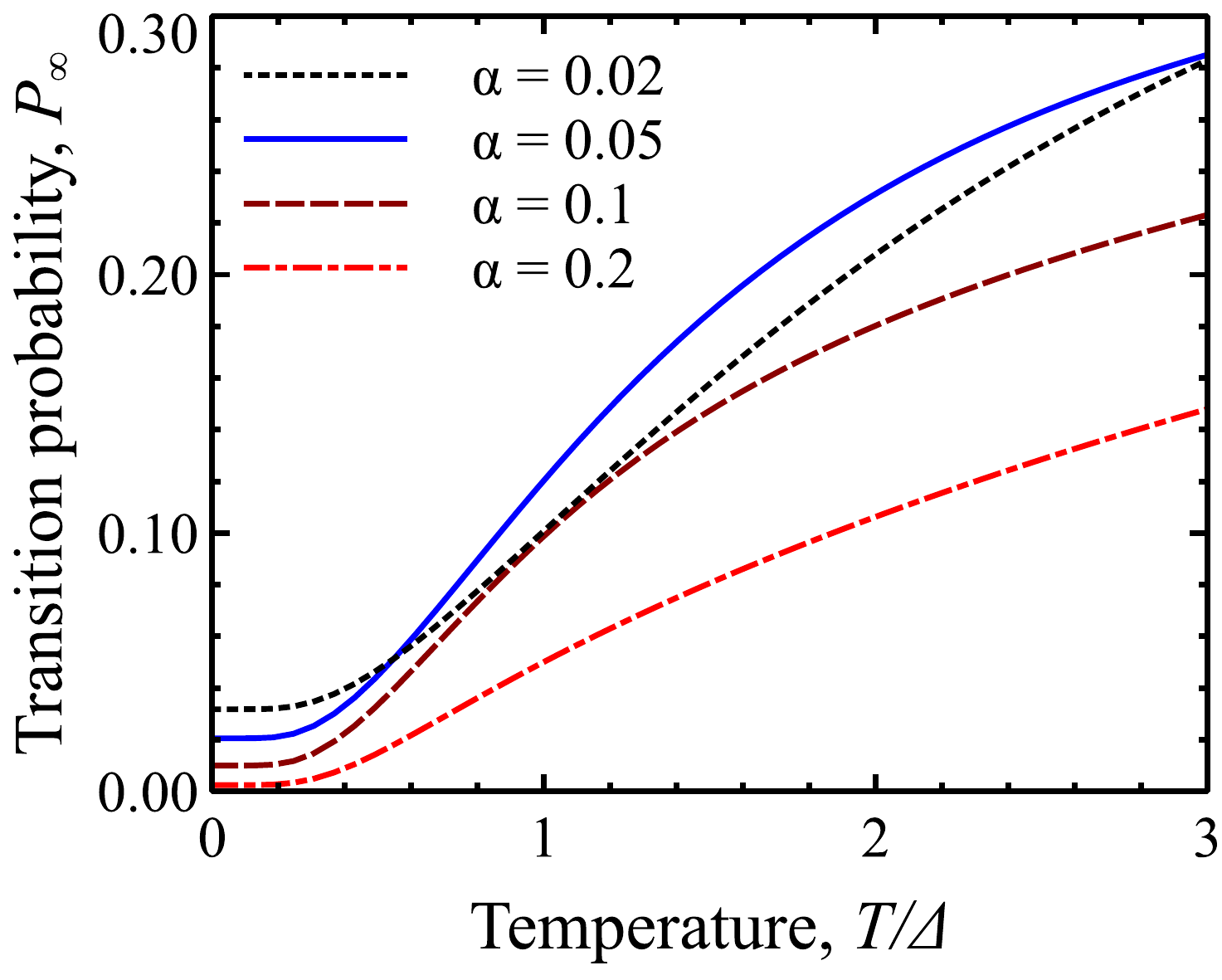}
\end{centering}
\caption{\label{fig:Pinfty_T}(Color online) Transition probability $P_\infty$ as a function of environment temperature $T$, at different values of coupling between the qubit and the environment for $\bm{n}\|\hat{\bm{z}}$.  Level-crossing speed $v=0.5\Delta^2$, the high energy cutoff for the environment is $E_c=10\Delta$ and $J_0=0$.  We take $P_\infty=P_{e}(t=4E_c/v)$. 
}
\end{figure}

From the above analysis, we conclude that a finite temperature of the environment leads to the ``equilibration'' between the ground and excited states of the qubit, and as temperature increases, 
the probability of the transition to the excited state in the LZ process increases monotonically, cf. Refs.~\cite{Ao1991,Grifoni1998}. This behavior is demonstrated in Fig.~\ref{fig:Pinfty_T}, where $P_\infty$ is shown as a function of $T$ for several values $\alpha$ of coupling between the qubit and its environment. We also note that the temperature effects appear at $T\gtrsim \Delta$, at smaller $T$, values of $P_\infty$ are characterized by the excitation through unitary evolution with the subsequent relaxation.

\begin{figure}
\begin{centering}
\includegraphics[width=0.9\columnwidth]{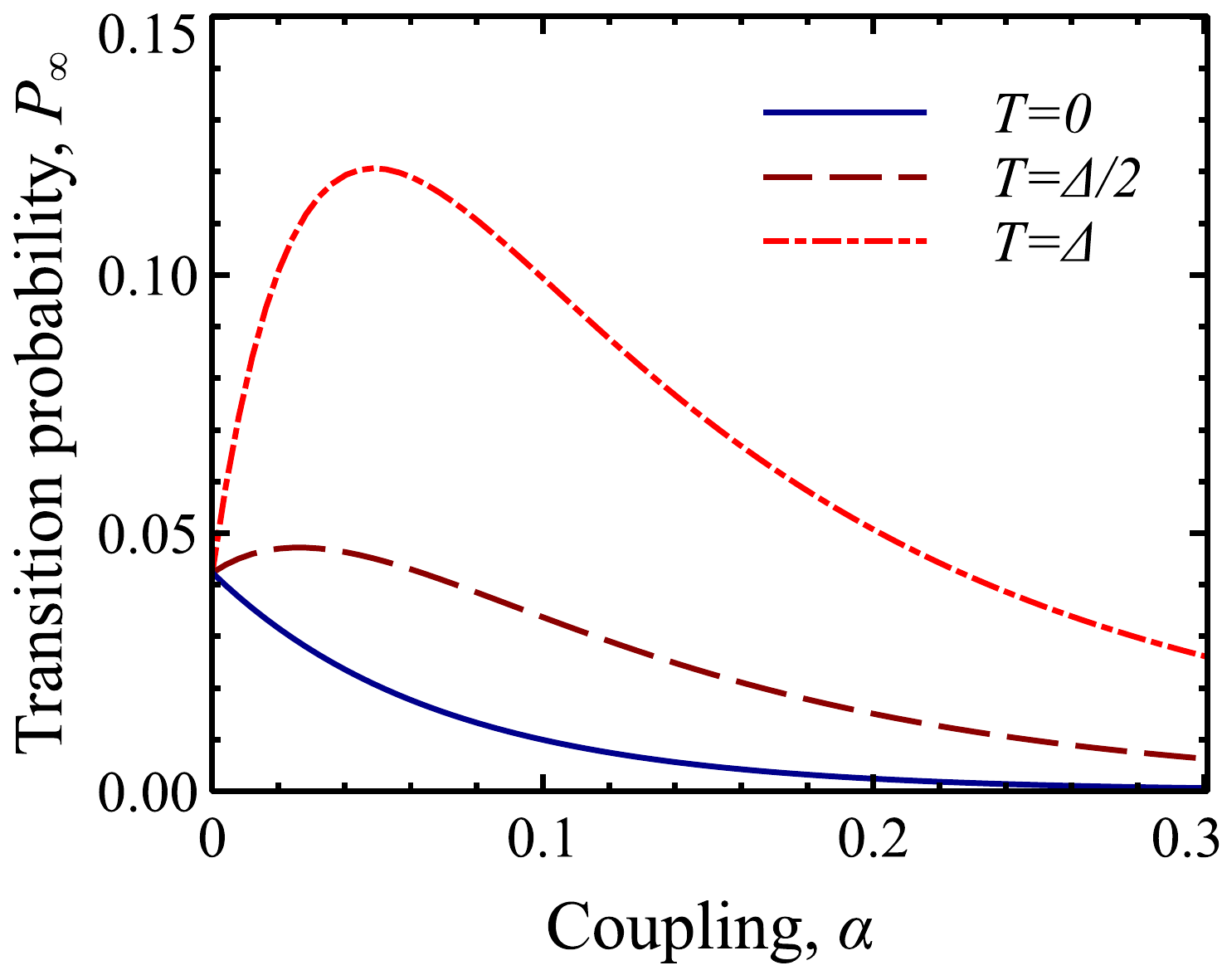}
\end{centering}
\caption{\label{fig:Pinfty_alpha}(Color online) Transition probability $P_\infty$ as a function of the coupling parameter of the qubit and the environment, $\alpha$, at different environment temperatures for $\bm{n}\|\hat{\bm{z}}$.  Level-crossing speed $v=0.5\Delta^2$, the high energy cutoff for the environment is $E_c=10\Delta$  and $J_0=0$.  We take $P_\infty=P_{e}(t=3E_c/v)$. 
}
\label{fig:Pinfty_vs_alpha}
\end{figure}

When we consider $P_\infty$ as a function of coupling $\alpha$ for several values of $T$, we observe a more complicated behavior.  For $T=0$, shown by the solid line in Fig.~\ref{fig:Pinfty_alpha}, the transition probability $P_\infty$ monotonically decreases from its value $P^{LZ}_\infty$, Eq.~\eqref{eq:LZideal}, as $\alpha$ increases, in agreement with Eq.~\eqref{eq:PinftyT0}.  At finite temperatures, $P_\infty$ increases for smaller values of $\alpha$, as the excitation process becomes more efficient and provides extra boost for transitions to the excited state in addition to that produced by unitary dynamics. However, this boost is only a linear function of $\alpha$, see Eqs.~\eqref{eq:finiteTlow} and \eqref{eq:finiteThigh}, and at stronger values of $\alpha$ the exponential dependence of $\Pi$ on $\alpha$ results in decreasing $P_\infty$ as $\alpha$ increases.

\begin{figure}[h]
\begin{centering}
\includegraphics[width=0.9\columnwidth]{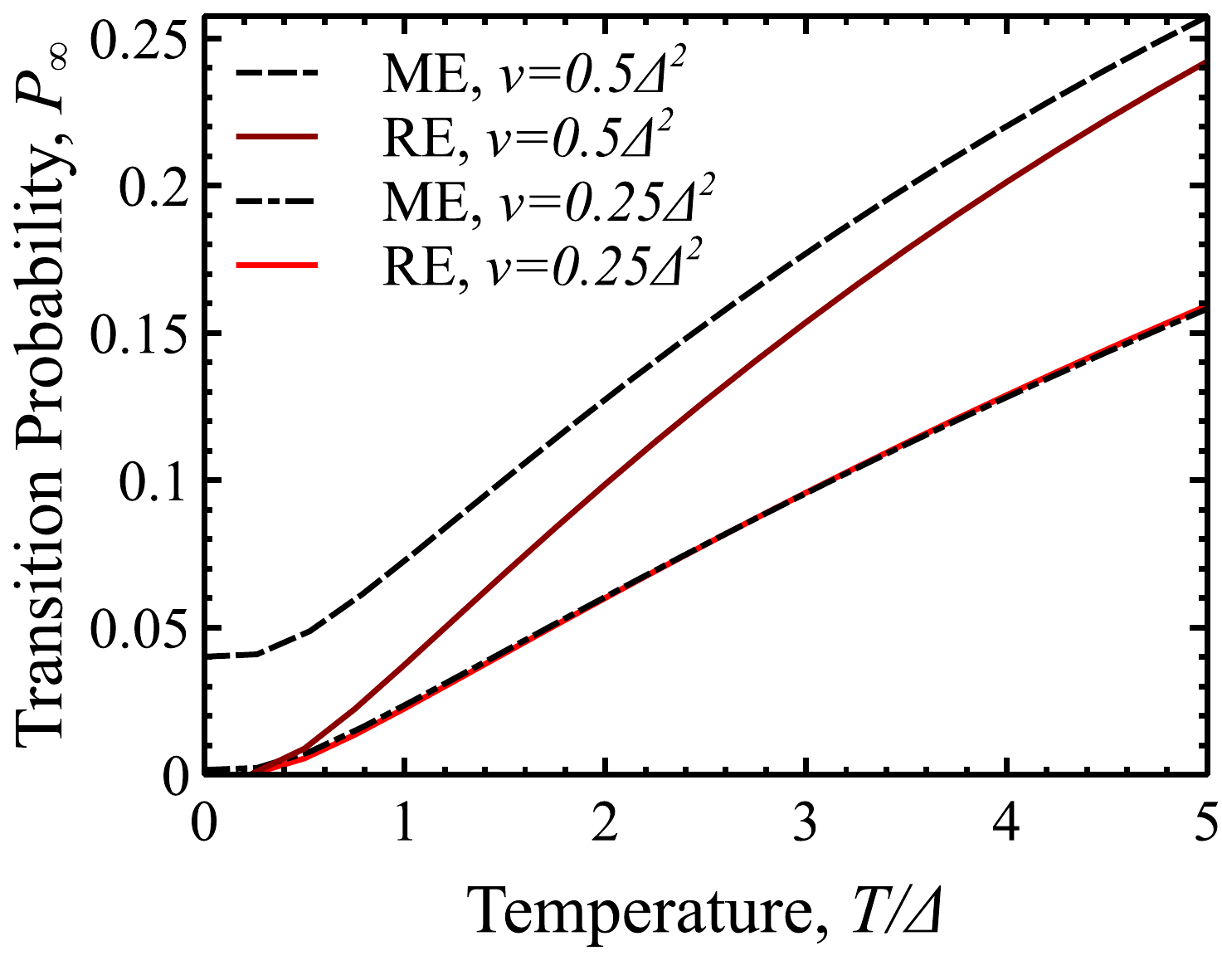}
\end{centering}
\caption{
(Color online) Transition probability $P_\infty$ as a function of environment temperature $T$ for $\bm{n}\| \hat{\bm{b}}$, at different values of drive velocity.  The high energy cutoff for the environment is $E_c=10\Delta$ and $J_0=0$. The solid lines represent solutions of rate equations Eq.~\eqref{eq:dmdt}. We take $P_\infty=P_{e}(t=4E_c/v)$. 
}
\label{fig:Pinfty_vs_T_long}
\end{figure}

\subsection{Longitudinal coupling }

We also consider the environment that produces fluctuating field along the direction of the control field, $\bm{n}\| \bm{b}$, in the Landau--Zener problem.  The decoherence rates in the BR equations~\eqref{eq:BRLZ} are given by
 \begin{subequations}
\label{eq:GammarLZlong}
\begin{align}
\Gamma_{r} & = \frac{\sin^2\eta }{2}J(W(t)) [N(W(t))+1],  
\\
\Gamma_{e} & = \frac{\sin^2\eta }{2}J(W(t))  N(W(t)),\\
\Gamma_{2} & = \frac{\Gamma_{r}+\Gamma_{e}}{2}
+
J_0 \cos^{2}\eta.
\end{align}
\end{subequations}
For this configuration of coupling between the qubit and environment, the matrix elements for transitions between different eigenstates of the qubit caused by the environment are small and 
the qubit flip rates $\Gamma_{r,e}$ are proportional to $\sin^2\eta \lesssim v^2\Delta^2/E^6(t)\leq v^2/\Delta^4$ and vanish fast for $|t|\gtrsim \Delta/v$ as $\Gamma_{r,e}\sim \Delta^2/v^4 t^6$.  Such fast decrease of the qubit flip rates in time simplifies either numerical or analytical integration of the BR equation and makes $P_\infty$ independent from the high-energy cutoff $E_c$.

In particular, for finite temperatures, when the BR equations can be reduced to the rate equations, time evolution  of  $m(t)=P_g(t)-P_e(t)$ is given by Eq.~\eqref{eq:dmdt} with $G_{LZ}(t)=\sin^2\eta$. The general solution of the rate equation takes similar form to Eq.~\eqref{eq:finiteT}:
\begin{equation}\label{eq:gensol_long}
\begin{split}
P_{\infty}&=\int_{-\infty}^{\infty}\Gamma_{e}(t)e^{-\int_{t}^{\infty}\Gamma_{l}(t')\coth\frac{W(t')}{2T}dt'}dt,\\
\Gamma_l&= \pi\alpha W(t)\sin^2\eta(t).
\end{split}
\end{equation}
Performing time integration in Eq.~\eqref{eq:gensol_long} gives
for $T\ll\Delta$:
\begin{equation}\label{eq:finiteT_long}
P_{\infty}=
\alpha v\sqrt{\frac{\pi^{3}}{32T\Delta^{3}}}
\exp\left(-\frac{\Delta}{T}\right)
\exp\left(-\frac{2\pi \alpha v}{3\Delta^{2}}\right).
\end{equation}
For high temperatures, $T\gg\Delta$, we obtain (see Appendix B)
\begin{equation}
\label{eq:finiteT_longb}
P_\infty=
\frac{1}{2}\left[
1-\exp\left(-\frac{3 \pi^2}{4}\alpha \frac{T v}{\Delta^3} \right)\right].
\end{equation}
As we mentioned above, the results in  Eqs.~\eqref{eq:gensol_long} and ~\eqref{eq:finiteT_longb} are independent from the cutoff energy $E_c$. Equation~\eqref{eq:finiteT_long}  shows that $P_{\infty}$ vanishes in the low temperature limit, unless we take into account non-adiabatic unitary evolution of the quantum state in the LZ problem.  In the limit of high temperatures $T\gg \Delta$, but still weak coupling, $\alpha v T\ll\Delta^3$, we obtain the linear dependence of $P_{\infty}$ on $T$: 
\begin{equation}
\label{eq:rho11LZ_hT_LF}
P_\infty =  \frac{3 \pi^2}{8}\alpha \frac{T v}{\Delta^3}, 
\end{equation}
which follows from Eq.~\eqref{eq:finiteT_longb}.

Since simple form of $P_{\infty}$ can not be obtained in the intermediate temperature regime, we numerically calculate the solution of rate equation as well as that of Bloch-Redfield equation for comparison, see Fig.~\ref{fig:Pinfty_vs_T_long}. When the level-crossing speed $v$ is small enough, the transition is mainly due to thermalization at short times and energy relaxation at longer times. In this regime, the rate and BR equations are in a very good agreement, as demonstrated in Fig.~\ref{fig:Pinfty_vs_T_long} for $v=0.25\Delta^2$.  
However, as the level--crossing speed increases, the non-adiabatic unitary evolution also contributes to the transition to the excited state increasing the probability for a system to be in the excited state.  
Since the non-adiabatic unitary evolution is not incorporated in the rate equations, the equations underestimate the probability of the excitation in the LZ process, compare the solid and dashed  curves in Fig.~\ref{fig:Pinfty_vs_T_long} for $v=0.5\Delta^2$.   

\section{Lindblad dephasing evolution}
\label{sec5}

We compare the results obtained from the BR equations in the case of longitudinal coupling with the theory based on the Lindblad equation for pure dephasing operators.   For both problems, the qubit Hamiltonian can be parametrized by the control field $\bm{b}=E(t) \{\sin\theta,\ 0,\cos\theta\}$, where $E(t)$ is the magnitude of the control field equal to the qubit level separation.  The corresponding equation for the density matrix in the adiabatic basis  has the form:
\begin{subequations}
\label{eq:L-LZ}
\begin{equation}
\dot \rho = \frac{iE(t)}{2}[\sigma_z,\rho]+\frac{i\dot \theta}{2}[\sigma_y,\rho]+\frac{\gamma}{2} (\sigma_z\rho\sigma_z-\rho).
\end{equation}
In the component form the above equation is
\begin{align}
\label{eq:rdiagLF}
\dot\rho_{00} & =\frac{\dot \theta}{2}(\rho_{01}+\rho_{10}),\quad 
\dot\rho_{11}=-\frac{\dot \theta}{2}(\rho_{01}+\rho_{10}),\\
\dot \rho_{01} & = (iE(t) -\gamma) \rho_{01} - \frac{\dot \theta}{2}(\rho_{00}-\rho_{11}),\\
\dot \rho_{10} & = (-iE(t) -\gamma) \rho_{10} - \frac{\dot \theta}{2}(\rho_{00}-\rho_{11}).
\end{align}
\end{subequations}
These equations are similar to Eqs.~\eqref{eq:BRall}, but because they are not written in the eigenstate basis, the last two equations contain extra terms.  Time derivatives of diagonal terms contain the off-diagonal terms of the density matrix multiplied by the quantity characterizing the off-diagonal part of the Hamiltonian, $\dot{\theta}$.  Time derivatives of the off-diagonal components of the density matrix have the terms identical to those in Eqs.~\eqref{eq:BRall} and the extra terms characterized by the diagonal matrix elements and parameter $\dot{\theta}$.  In this section we again consider the two cases: (1) the qubit rotation with a constant angular velocity $\dot \theta=\Omega$, i.e. $\theta(t) = \Omega t$, and $E(t) = \Delta$; (2) the LZ problem with $E(t) = \sqrt{\Delta^2+v^2t^2}$ and  $\theta(t) = \arctan \Delta/vt$.

\subsection{Rotating field}

\begin{figure}

\begin{centering}
\includegraphics[width=0.9\columnwidth]{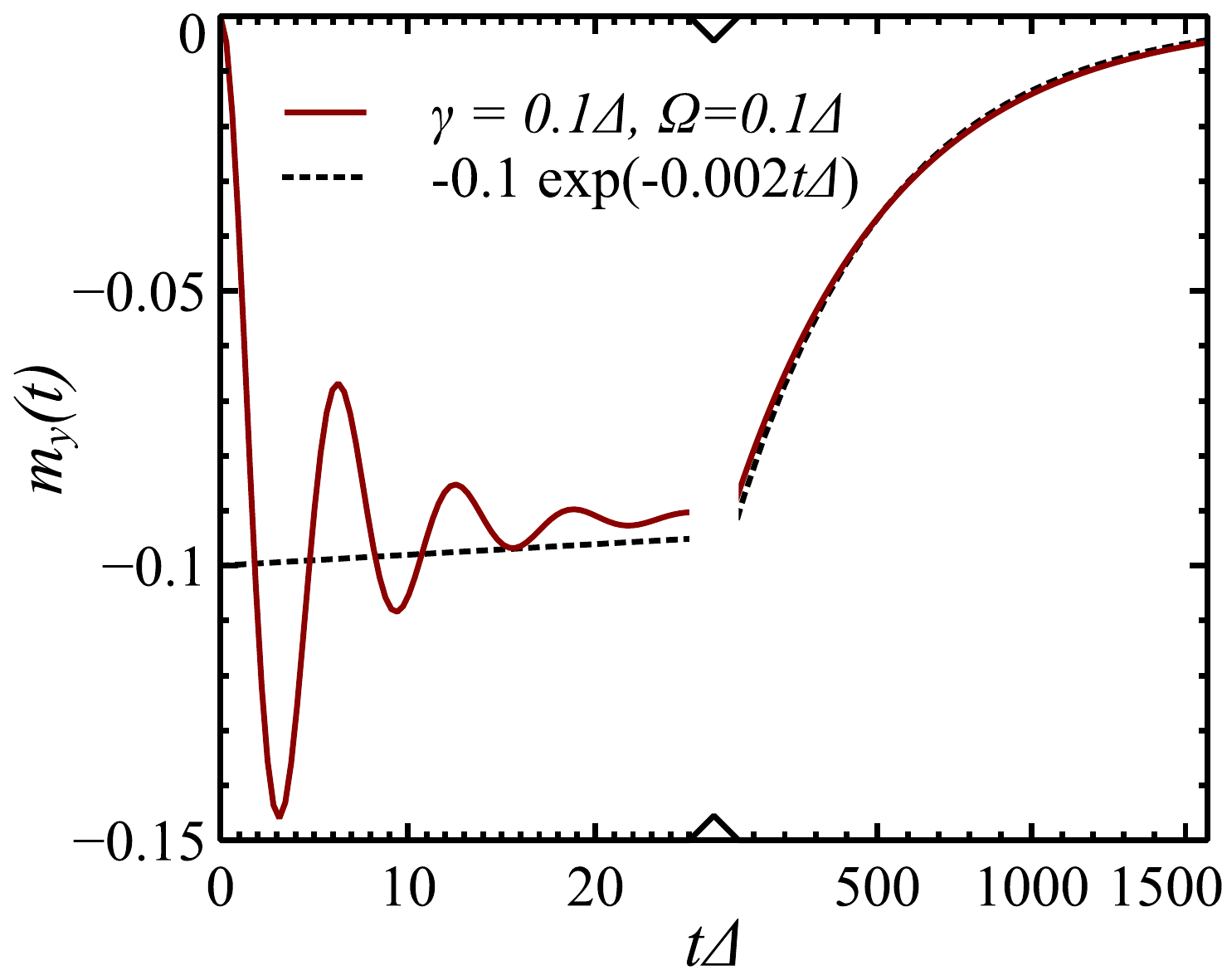}
\end{centering}
\caption{\label{fig:my-L}(Color online) Polarization $m_y(t)$ as a function of time $t$ for dephasing Lindblad evolution.  The decoherence rate $\gamma=0.1\Delta$ and rotation velocity $\Omega=0.1\Delta$.  After the rotation starts, polarization shows an oscillatory behavior originating from the qubit precession, at longer times the precession stops and the qubit relaxes to unpolarized state according to Eq.~\eqref{eq:myL}.
}
\end{figure}

When the control field rotates in $(x-z)$ plane, $\bm{b}(t)=\Delta\{\sin\Omega t,\ 0,\cos\Omega t\}$, the effective Hamiltonian is time independent. To make a comparison with the calculation of BR equations, one can look for a quasi-stationary state solution of the density matrix at time scale $t\sim 1/\gamma$ with ansatz that the off-diagonal elements are  $\rho_{01/10}\propto \Omega$. We disregard $\Omega^2$ terms for $\dot \rho_{00/11}$ and take $\rho_{00}=1$.  Then, we have $\rho_{01}=\Omega/2(i\Delta-\gamma)$,   $\rho_{10}=\Omega/2(-i\Delta-\gamma)$ and the out of plane qubit projection is\cite{Avron2011} 
\begin{equation}
m_y{(L)}= - \frac{\Omega}{2}\frac{\Delta}{\Delta^2+\gamma^2}.
\end{equation}
We argue, however, that the above expression does not hold for authentic steady state, $\dot{\hat{\rho}}=0$, at longer times and for general configuration of the initial conditions. We present the result of numerical integration of the Lindblad equations~\eqref{eq:L-LZ} in Fig.~\ref{fig:my-L} for $\Omega=0.1\Delta$ and $\gamma=0.1\Delta$.  In our calculation, we consider the case when the qubit is prepared in the ground state prior to rotation for $t<0$.  When the rotation starts, the Hamiltonian acquires extra terms $\sim \Omega$ and the qubit exhibits a precession around new direction of the control field.  This precession is reduced by the decoherence with rate $\Gamma_2\simeq \gamma $  and the oscillatory component in $m_y(t)$ vanishes for times $t\sim 1/\gamma$.  

At longer times, the diagonal matrix elements start changing as well and the system will eventually relax to $\rho_{00}=\rho_{11}=1/2$ and 
$\rho_{01}=\rho_{10}=0$.   The reason for this behavior is that at long times, the diagonal elements acquire significant changes even though these changes have small factor  $\Omega^2$.  In the language of the BR equation, the Lindblad pure dephasing operator contains relaxation and excitation components in the eigenstate basis of the transformed Hamiltonian $\hat{H}_0^V$ and $\Gamma_e=\Gamma_r=\gamma \Omega^2/(\Delta^2+\Omega^2)$, which is the high temperature limit because it does not distinguish processes with absorption or emission of environment excitations. Correspondingly, the density matrix reaches the high-temperature limit with equal probabilities of occupation of eigenstates of the qubit Hamiltonian
\begin{equation}
\label{eq:myL}
m_y^{(L)}(t)= -\frac{\Omega}{\sqrt{\Delta^2+\Omega^2}}\exp\left(-\frac{2\Omega^2 \gamma t }{\Omega^2+\Delta^2}\right).
\end{equation}
This asymptotic behavior is consistent with the result obtained from the numerical solution of the Lindblad equation~\eqref{eq:L-LZ}, shown in   Fig.~\ref{fig:my-L}.

\subsection{Landau--Zener problem}

\begin{figure}
\begin{centering}
\includegraphics[width=0.9\columnwidth]{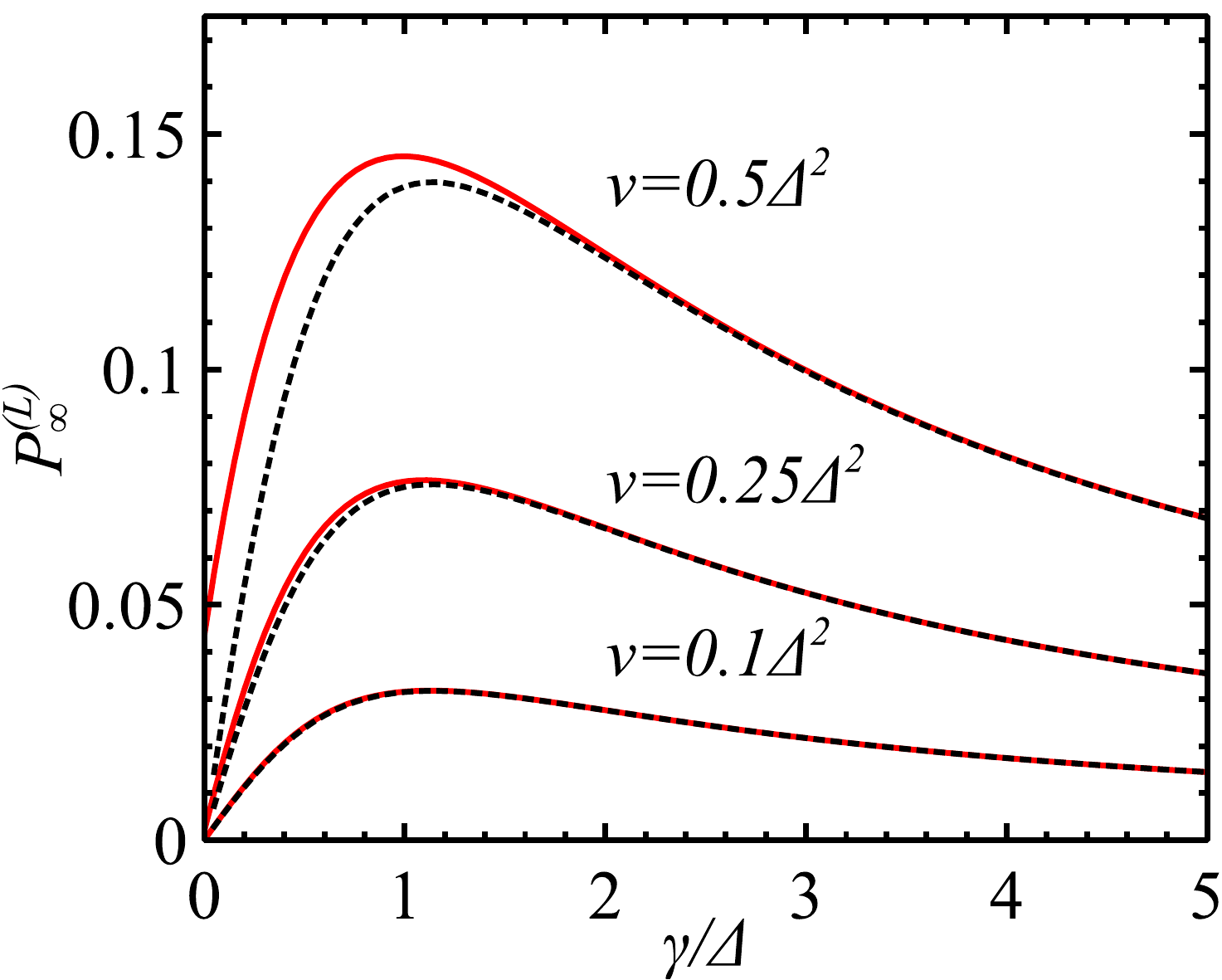}
\end{centering}
\caption{\label{fig:L-LZ}(Color online) Transition probability $P_\infty$ as a function of daphasing rate $\gamma$ for different level--crossing speeds $v/\Delta^2 =0.1,\ 0.25,\ 0.5$. Solid lines are numerical solution of the Lindblad equation, Eq.~\eqref{eq:L-LZ}, and dashed lines are given by Eq.~\eqref{eq:PLinfty}.
}
\end{figure}

The expression for Landau--Zener problem to the lowest order in $v$ can be obtained from the explicit form of the Lindblad equation~\eqref{eq:L-LZ} with $E(t)$ given by Eq.~\eqref{eq:H0U1LZ} and $\Omega=v\Delta/E^2(t)$.  We assume that the changes in the system are slow and disregard $\dot\rho_{01}$ and $\dot\rho_{10}$ in Eqs.~\eqref{eq:L-LZ}. 
Then we find $\rho_{01}=\Omega/2(-iE+\gamma)[\rho_{00}-\rho_{11}]$ and $\rho_{10}=[\rho_{01}]^*$.  Substituting these expressions to Eq.~\eqref{eq:rdiagLF}, we obtain:
\begin{align}
\label{eq:PLinfty}
P_\infty^{(L)} & =\frac{1}{2}\left[1-\exp\left(-\gamma \int_{-\infty}^{\infty}
\frac{v^2\Delta^2}{E^4(t)}\frac{dt}{\gamma^2+E^2(t)}\right)\right]
\nonumber
\\
 & = \frac{1}{2}\left[1-\exp\left(-\frac{\pi v}{2\Delta^2}R\left(\frac{\gamma}{\Delta}\right)\right)\right],
\end{align}
where 
\begin{equation}
R(x)=\frac{2+(x^2-2)\sqrt{x^2+1}}{x^3\sqrt{x^2+1}}.
\end{equation}
In the limit $v\ll \Delta^2$, we recover the result of Ref.~\cite{Avron2010}:
\begin{align}
P_\infty^{(L)}  & =-\frac{\pi v}{4\Delta^2}R\left(\frac{\gamma}{\Delta}\right).
\end{align} 

At small decoherence rate and slow drive, $\gamma\ll\Delta$, we take $R(x\ll 1)\simeq 3x/4$ and reproduce the previous result, Eq.~\eqref{eq:rho11LZ_hT_LF}, if we identify $\gamma=2\pi \alpha T$.  The agreement between Eqs.~\eqref{eq:rho11LZ_hT_LF} and \eqref{eq:PLinfty} has a simple interpretation.  The Lindblad equation can be viewed as the high--temperature limit of the BR equation for the Ohmic 
environment\cite{Whitney2008}.  The Lindblad equation~\eqref{eq:L-LZ} is written in the basis that does not completely diagonalize the Hamiltonian operator, and when we rewrite 
this equation in the basis diagonalizing matrix $E(t)\hat{\sigma}_z+\Omega\hat{\sigma}_x$, we arrive to the collapse operators that represent transition processes between the eigenstates with equal excitation and relaxation rates  
$\Gamma_{e,r}^{(L)}\approx \gamma (\Omega^2(t)/E^2(t))$.  It is the excitation processes that cause transitions of the system to the excited state with  the population of an excited state $P_\infty$ in accordance with Eq.~\eqref{eq:PLinfty}.  To account for finite temperatures, the Lindbladian operators are to be written in the eigenstate basis of the ``dressed'' Hamiltonian, see Ref.~\cite{Shevchenko2013}. 

Large decoherence rate, $\gamma\gg \Delta$, suppresses the off-diagonal elements of the density matrix, and effectively reduces the excitation and relaxation rates $\sim \gamma \Omega^2/(E^2+\gamma^2)$. 
As a result, the qubit is more likely to stay in its ground state without experiencing an excitation during the LZ avoided level crossing. The maximum of $R(x_m)\approx 0.42$ is reached at $x_m=1.14$.

We compare Eq.~\eqref{eq:PLinfty} (dashed lines) with the result of numerical integration of the Lindblad equation~\eqref{eq:L-LZ} (solid lines) in Fig.~\ref{fig:L-LZ}. We observe that at stronger decoherence rate, when the off-diagonal unitary terms in the evolution of the density matrix can be neglected in comparison with the decoherence terms, $\gamma\gg v/\Delta$ in the LZ problem, the two solutions are equivalent.

\section{Discussion and Conclusions}
\label{sec6}
In conclusion, we have presented a  detailed analysis of the dynamics of an open quantum system in the presence of time-varying control field. Specifically, we applied the Bloch-Redfield formalism to a spin-1/2 system whose Hamiltonian varies slowly with time and investigated two problems. In the first problem, we studied the response of a qubit to a rotating control field of the qubit with a fixed magnitude.  We noted that when the qubit basis is transformed to keep the effective Hamiltonian in the diagonal form, which is required for proper perturbative analysis of the coupling between the qubit and its environment, the transformed Hamiltonian acquires extra gauge terms. The gauge terms result in the modification of the qubit--environment coupling and are related to the renormalization of the mass and friction terms due to changing parameters of the Hamiltonian, cf. Ref.~\cite{D'Alessio2013a}.  The exact form of the renormalization depends on a particular orientation of the control field with respect to the fluctuating environment field. We have illustrated this scenario by considering different orientations of the environment field: (1) control field and fluctuations are always perpendicular to each other, and the corresponding relaxation rates are time-independent; (2) control and fluctuation fields are parallel only at some moments of time, in which case the relaxation rates significantly oscillate in time; (3) fluctuations are always along the direction of the control field, then the relaxation rates are small in the parameter given by the ratio of the rotation velocity and level separation.

Our analysis offers a clear evidence of robustness of topological features against external noises. To see this one needs to consider a long time limit where the qubit density matrix reaches a steady state solution that at zero temperature coincides with the ground state of the effective Hamiltonian.  When this ground-state qubit configuration is looked at  in the original laboratory basis, the qubit has a constant projection in the direction perpendicular to the plane of rotation and the magnitude of the projection is proportional to the product of rotation velocity of the control field and the Berry curvature of the qubit ground state. In the long time limit, this response is unaffected by the environmental coupling field, at least for zero temperature environment. This relation of the response at long times and the Berry curvature can be utilized as a practical method for measurements of the Chern number\cite{Xiao2010} of a quantum system.

We  also considered an environment with a very sharp spectral function.  We represent this environment by a quantum harmonic oscillator that has internal relaxation.  In this case we solve the Lindblad master equation for the system of coupled qubit and oscillator and find that the results are qualitatively similar to the solution of the BR equation with properly chosen relaxation rates.  

In the second example, we revisited the Landau--Zener problem.  In this case, the modification of the matrix elements for transitions between eigenstates of the qubit Hamiltonian is essential, even though it was not always taken into account.\cite{Wubs2006,Saito2007}  The eigenstate basis that is necessary to use in treatment of interaction of the qubit with its environment is also convenient for numerical evaluation because in this basis the system behavior   
during the Landau--Zener level crossing is represented by a smooth function that quickly reaches its long-time asymptotic value.  

For a qubit weakly coupled to the environment, the evolution, long after the level crossing, reduces to suppression of the off-diagonal elements of the density matrix and relaxation of the excited state to the ground state, the latter is accurately described by the rate equations.  For the fluctuating field along the asymptotic direction of the control field, the relaxation rate decreases as the level separation increases due to suppression of the matrix elements of qubit transition between eigenstates caused by the environment.  However, this suppression is not sufficient to cut the relaxation in the long time limit, and the relaxation results  in a power law decay of the excited state, until the separation between the qubit states exceeds the ultra-violet cutoff of the environment. 

At finite temperature, in addition to enhancement of decoherence rates for the qubit, the excitation processes produce transitions from the ground to the excited qubit states, eventually increasing the probability for the qubit to appear in the excited state after the transition.  The BR equations accurately describe the crossover for the Landau--Zener transition in an isolated quantum system, Eq.~\eqref{eq:LZideal}, with unitary evolution, to the open system at arbitrary temperature, see Sec.~\ref{sec4}.  

Furthermore, we compare the results obtained from the generalized BR equations with that from the Lindblad master equation.  In particular, we focused on the case of pure dephasing Lindblad superoperators,\cite{Avron2010,Avron2011} that are equivalent to the longitudinal coupling of the environment (fluctuating field of the environment is along the control field).  We found that the two results are consistent in the high temperature limit, when the Lindblad and BR equations are equivalent, but application of the Lindblad equation for a system coupled to low temperature environment may result in unphysical solutions.

Finally, we note that  the generalization of the Bloch--Redfield equations can be applied to accurately evaluate the fidelity of quantum gates.  By taking into account proper modification of the transition and dephasing rates caused by time-varying parameters in the Hamiltonian, optimization techniques for gate operations can be further improved.  Similarly, the BR equations for time-dependent Hamiltonian are also required for accurate description of protocols for adiabatic quantum computing and  the Berry phase measurement in recent experiments.~\cite{Berger2013}  

\acknowledgements
We thank I. Aleiner, A. Glaudell, F. Nori, A. Polkovnikov, S. Shevchenko and A. Levchenko
for fruitful discussions.  The work was supported by NSF Grants No. DMR-1105178 and DMR-0955500, ARO and LPS Grant No. W911NF-11-1-0030.

\appendix
\section{Solution of rate equations for the avoided level crossing}\label{App:A}
Here we evaluate the integral in Eq.~\eqref{eq:finiteT}.  Notice that 
while the integral over $t'$ in the exponent, 
\begin{equation}
\label{eq:A1}
I_1(t)=\int_{t}^{\infty}\Gamma_{0}(t')\coth\frac{W(t')}{2T}dt'
\end{equation}
originates on long interval from $\sim \Delta/v$ to $E_c/v$, the second integral converges for time $|t|\lesssim T/v$, for not very large temperatures, we can replace the low limit of integration in Eq.~\eqref{eq:A1} by $t=0$.  In this case, we have
\begin{equation}
\label{eq:A2}
P_{\infty}=e^{-I_1(0)} I_2,\quad I_2=\int_{-\infty}^{\infty}\Gamma_{e}(t)dt,
\end{equation}
where $W(t)=\sqrt{\Delta^{2}+v^{2}t^{2}+v^{2}\Delta^{2}/(\Delta^{2}+v^{2}t^{2})^{2}}\simeq\sqrt{\Delta^{2}+v^{2}t^{2}}$,
$\Gamma_{e}(t)=G_{LZ}(t)J(W(t))N(W(t))/2$, $\Gamma_{0}=G_{LZ}(t)J(W(t))/2$
with $G_{LZ}\simeq\Delta^{2}/(\Delta^{2}+v^{2}t^{2})$ and $J(\omega)=2\pi\alpha\omega\exp(-\omega/Ec)$.
First, let us change the integration variable $t=\sqrt{s^{2}-\Delta^{2}}/v$
such that $dt=s/v\sqrt{s^{2}-\Delta^{2}}ds$ and the
integral in the exponential then reads
\begin{equation}
I_{1}(0)=\int_{\Delta}^{\infty}\frac{\pi\alpha\Delta^{2}}{v\sqrt{s^{2}-\Delta^{2}}}\coth\frac{s}{2T}\exp(-s/E_{c})ds.
\end{equation}
This integral can be evaluated in two cases. First, we consider the
low temperature limit $T\rightarrow0$, in which the hyperbolic cotangent
$\coth s/2T\rightarrow1+2\exp(-s/T)$. Therefore, the integral is
obtained
\begin{eqnarray}
I_{1}(0) & = & \frac{\pi\alpha\Delta^{2}}{v}\left[2K_{0}(\Delta/T)+K_{0}(\Delta/E_{c}))\right],
\end{eqnarray}
where $K_{0}(x)$ is the 0th order modified Bessel function of the second
kind with the following asymptotes: $K_0(x)\simeq \sqrt{\pi/2x}\exp(-x)$ for $x\gg 1 $and $K_0(x)\simeq -\ln (x e^\gamma/2)$ for $x\ll 1$, $\gamma\simeq 0.577$ is the Euler constant.  As the result, for $T\ll \Delta$, we have 
\begin{equation}
I_1(0)\simeq 
\frac{\pi\alpha\Delta^{2}}{v}\left[\sqrt{\frac{2\pi T}{\Delta}}e^{-\Delta/T}+\ln(2E_{c}/\Delta)-\gamma\right].
\end{equation}
The first term can be disregarded for $T\ll \Delta$.

At higher temperatures, there is a stronger contribution to $I_1(0)$ originating from short time interval $|t|\lesssim T/v$.  We can estimate this contribution as 
\begin{eqnarray}
\delta I_{1} & = & \frac{\pi\alpha\Delta^{2}}{v}\int_{\Delta}^{\infty}
\frac{2T}{s\sqrt{s^{2}-\Delta^{2}}}ds
=\frac{\pi\alpha\Delta^{2}}{v}\frac{\pi T}{\Delta}.
\end{eqnarray}
We emphasize that this is the contribution which we do not evaluate correctly when replace Eq.~\eqref{eq:finiteT} by Eq.~\eqref{eq:A2}.  Therefore, we can treat the above expression for $\delta I_1$ as the boundary of applicability of our approximation, indicating that transition from Eq.~\eqref{eq:finiteT} to~\eqref{eq:A2} is justified not for very high temperatures, such that $\delta I_1\ll 1$.

Next, we evaluate the integral 
\begin{equation}
I_{2}=\int_{-\infty}^{\infty}\Gamma_{e}(t)dt=2\int_{\Delta}^{\infty}ds\frac{2\pi\alpha\Delta^{2}}{v\sqrt{s^{2}-\Delta^{2}}}\frac{\exp(-s/E_{c})}{\exp(s/T)-1}.
\end{equation}
As before, we first consider the low temperature limit, $T\ll \Delta$, in which we approximate $1/[\exp(s/T)-1]\simeq \exp(-s/T)$.
Then the integral becomes 
\begin{equation}
I_{2}\simeq 
\frac{2\pi\alpha\Delta^2}{v}K_{0}(\Delta /T)\simeq 
\frac{2\pi\alpha\Delta^2}{v}\sqrt{\frac{\pi T}{2\Delta}}e^{-\Delta/T}.
\end{equation}
In the high temperature limit, we utilize $1/[\exp(s/T)-1]\simeq T/s$, and we obtain
\begin{equation}
I_2= \frac{\pi^2 \alpha T\Delta}{v}.
\end{equation}
This equation is valid for high temperature limit $T\gg \Delta$, provided that our substitution of  Eq.~\eqref{eq:finiteT} by~\eqref{eq:A2} is justified, or 
$\alpha T\ll v/\Delta$.

To sum up, we evaluated $P_{\infty}$ in the limits of low and moderately high temperatures. The results are presented by Eqs.~\eqref{eq:finiteTlow} and \eqref{eq:finiteThigh}.

\section{Solution of rate equations for the avoided level crossing for environment with longitudinal coupling} 

For the longitudinal coupling, the transition probability $P_\infty$ in  limit of low temperatures $T\lesssim \Delta$ can be evaluated similarly to the calculations in Appendix A.  
We replace Eq.~\eqref{eq:gensol_long}, where the integral over time $t$ converges fast for $|t|\lesssim T/v$, by the following expression
\begin{align}
P_{\infty} & = I_2 e^{-I_1},\quad 
I_2 = \int_{-\infty}^{\infty} \Gamma_{e}(t) dt \\
I_1 & = 
\int_{0}^{\infty}\Gamma_{l}(t) dt,
\end{align}
where in the last integral we take the lower limit of integration to zero and $\coth W/2T \to 1$.
In the above expression, $W(t)=\sqrt{\Delta^{2}+v^{2}t^{2}+v^{2}\Delta^{2}/(\Delta^{2}+v^{2}t^{2})^{2}}\simeq\sqrt{\Delta^{2}+v^{2}t^{2}}$,
$\Gamma_{e}(t)=G_{LZ}(t)J(W(t))N(W(t))/2$, $\Gamma_{0}=G_{LZ}(t)J(W(t))/2$
with $G_{LZ}\simeq v^{2}\Delta^{2}/(v^{2}t^{2}+\Delta^{2})^{3}$ and
$J(\omega)=2\pi\alpha\omega\exp(-\omega/Ec)$. Similarly, let us change
the integration variable $t=\sqrt{s^{2}-\Delta^{2}}/v$ such that
$dt=s/v\sqrt{s^{2}-\Delta^{2}}ds$. 
The integral $I_1$ then reads
\begin{equation}
I_{1}=\int_{\Delta}^{\infty}\frac{\pi\alpha\Delta^{2}v}{s^{4}\sqrt{s^{2}-\Delta^{2}}}ds =\frac{2\pi\alpha v}{3\Delta^{2}}.
\end{equation}
We note that this integral converges fast and the high-energy cutoff of the environment can be omitted. 
Similarly, the integral over $\Gamma_{e}(t)$ can be rewritten as
\begin{equation}
\begin{split}
I_{2} = & \int_{\Delta}^{\infty}\frac{2\pi\alpha\Delta^{2}v}{s^{4}\sqrt{s^{2}-\Delta^{2}}}\frac{ds}{\exp\left(s/T\right)-1}\\
 & \simeq \int_{\Delta}^{\infty}\frac{\sqrt{2}\pi\alpha\Delta^{3/2}v}{s^{4}\sqrt{s-\Delta}}\exp\left(-s/T\right)ds\\
 & \simeq \alpha v\sqrt{\frac{\pi^{3}}{32T\Delta^{3}}}\exp\left(-\Delta/T\right).
\end{split}
\end{equation}

In the high temperature limit, we follow a different approach.
 We assume that the environment is at high temperature and the relaxation rates are enhanced by factor $T/W(t)$.  In this case, we also have a fast convergence of integrals $\int \Gamma_0 (t) dt$ at $|t|\lesssim \Delta/v$ and for $T\gg \Delta$, we can simplify the rate equation~\eqref{eq:dmdt} to
\begin{equation}
\frac{dm}{dt}  = -  2\pi\alpha T \frac{v^2\Delta^2 }{W^2(t)E^4(t)}  m(t). 
\end{equation} 
This equation can be integrated to find $m(t)$ with initial condition $m(-\infty)=1$, and used to define $P_\infty=(1-m(+\infty))/2$: 
\begin{align}
\label{eq:rho11LZ_hT_LF0}
P_\infty & =
\frac{1-e^{-I_3}}{2}
,\quad
I_3  = 2\pi\alpha T\int_{-\infty}^{\infty}  \frac{v^2\Delta^2 d t}{W^2(t)E^4(t)}.
\end{align}
For $v\ll \Delta^2$, we obtain 
\begin{align}
\label{eq:rho11LZ_hT_LF1}
I_3= \frac{3 \pi^2}{4}\frac{\alpha T v}{\Delta^3} ,
\end{align}
arriving to Eq.~\eqref{eq:finiteT_longb}.

% \bibliography{adiabatic_process-2.bib}

\end{document}